\documentclass[a4paper,11pt]{article}
\usepackage{pos}
\usepackage{todonotes}

\renewcommand{\logo}{\relax}
\newcommand*\aap{A\&A}

\newcommand*\actaa{Acta Astron.}
\newcommand*\aj{AJ}

\newcommand*\apj{ApJ}

\newcommand*\apjs{ApJS}

\newcommand*\mnras{MNRAS}

\newcommand*\nar{New A Rev.}

\newcommand*\pasj{PASJ}

\def\t{$\times$}
\newcommand{\tim}[1]{\ensuremath{\times 10^{#1}}}
\def\l{\ensuremath{\lambda}}

\newcommand{\Rsun}{R$_\odot$}
\newcommand{\Msun}{M$_\odot$}
\newcommand{\kms}{km\,s$^{-1}$}

\newcommand{\erg}{erg\,s$^{-1}$}

\newcommand{\Halpha} {H$\alpha$}
\newcommand{\Hbeta}  {H$\beta$}
\newcommand{\Hgamma} {H$\gamma$}
\newcommand{\Hdelta} {H$\delta$}

\newcommand{\HeI}  {He\,{\sc i}}
\newcommand{\HeII} {He\,{\sc ii}}

\title{The accretion discs in WZ Sge-type stars in deep quiescence. How do they outburst?}
\ShortTitle{Accretion discs in WZ Sge-type stars}

\author*[a]{Vitaly Neustroev}
\author[a]{Paula Kvist}
\author[a]{Miikka Siitonen}
\author[a]{Veera Vuolteenaho}

\affiliation[a]{Space Physics and Astronomy research unit, PO Box 3000, FIN-90014 University of Oulu, Finland}

\emailAdd{vitaly@neustroev.net}

\abstract{
WZ Sge-type stars are an extreme subclass of dwarf novae characterised by very rare, large-amplitude superoutbursts. Within
the disc instability model (DIM), such events are explained as being triggered by enhanced mass transfer from the donor star.
We present an analysis of observations of a sample of WZ Sge-type systems in deep quiescence to assess the consistency of DIM
predictions with their observed properties. We find that accretion discs in quiescent WZ Sge-type systems have very low
mass-accretion rates of a few \tim{-13} \Msun\ yr$^{-1}$. The discs are entirely optically thin, and their physical conditions
-- such as surface density and effective temperature -- remain well below the DIM thresholds required to trigger an outburst.
Observationally, no increase in disc brightness is detected prior to the superoutburst, indicating the absence of a transition
to an optically thick state, in contrast to DIM predictions of a gradual disc thickening preceding the instability. We therefore
find no observational evidence that superoutbursts in WZ Sge-type systems are triggered by enhanced mass transfer from the donor.
Furthermore, the inferred mass-transfer rates in these objects ($\dot{M}_{\rm tr}$$\sim$5\tim{-12} \Msun~yr$^{-1}$) are at least
an order of magnitude lower than commonly assumed. We argue that the widely adopted value of $\dot{M}_{\rm tr}$ for the prototype
object WZ~Sge is likely overestimated. Finally, we show that in quiescence the accretion disc radius in all systems is close to the
tidal truncation radius and exceeds the 3:1 resonance radius, confirming earlier results and calling into question the standard
interpretation of superhump formation.
}

\FullConference{87th Fujihara Seminar: The 50th Anniversary Workshop of the Disk Instability Model in Compact Binary Stars (DIM50TH2025)\\
22-26 September 2025\\
Tomakomai, Japan\\}

\begin{document}

\maketitle

\section{Introduction}

Cataclysmic variables (CVs) are interacting binary systems with a white dwarf (WD) as the primary component and a low-mass star
as the secondary component. The Roche-lobe filling secondary loses matter via the inner Lagrangian point to the primary. In the
absence of a strong magnetic field, the material transferred from the donor star forms an accretion disc around the WD and
progressively spirals down onto its surface \cite{Warner}. The mass transfer rate in CVs decreases as they evolve from longer
to shorter orbital periods \cite{kni11CVdonor}. When the mass-transfer rate falls below a critical threshold, the disc may
undergo thermal instability, leading to outbursts \cite{osa74DNmodel}. CVs which undergo such outbursts are named dwarf novae
(DNe). The disc instability model (DIM) is the most commonly used model to explain outbursts and their onset
(see \cite{las01DIDNXT,ham20CVreview} for review).

The DIM reproduces reasonably well many of the main characteristics of ordinary DNe. However, the DIM, in its basic form, was
unable to account for certain defining features observed in other DN subclasses, such as the Z~Cam- and SU~UMa-types of DNe.
The model must be enriched by including physical processes that may enable the reproduction of those objects' observational
properties \cite{sma00DNunsolved}. For example, in SU~UMa-type stars\footnote{SU UMa-type stars are short-period DNe
($P_{\rm orb}$$\lesssim$2~hr, with a few longer period examples) exhibiting two types of outbursts --- normal outbursts lasting
a few days, and superoutbursts which have a larger amplitude and a longer duration of a few weeks. The defining property of
superoutbursts is the presence of superhumps, low-amplitude modulations with a period of a few per cent longer than the orbital
one.}, the binaries with low mass ratios $q$ ($\equiv$$M_2$/$M_{\rm wd}$, where $M_2$ and $M_{\rm wd}$ are the masses of the
donor and WD, respectively), their superoutbusts and superhumps are due to a tidal-thermal instability (TTI), first proposed by
Osaki \cite{osa89suuma}. According to the TTI model, during a normal outburst, the accretion disc expands and then, during the
following quiescence period, contracts with time, settling into a slightly larger radius than before the outburst. After several
cycles, the disc radius reaches the 3:1 resonance radius that triggers the TTI, which, in combination with the standard
thermal-viscous instability, results in a superoutburst \cite{las01DIDNXT,bua02suumamodel,osa05DImodel,ham05XTdisk}.

However, despite its success in explaining observational features of superoutbursts in ordinary SU~UMa-type DNe, the DIM fails
to reproduce some of the fundamental properties of superoutbursts in WZ Sge-type stars. DNe of the WZ Sge type compose an
extreme subgroup of SU~UMa-type stars, which show very rare (approximately once a decade), large-amplitude (up to 9-10 mag)
superoutbursts lasting for 3--5 weeks, while normal (lower amplitude and shorter duration) outbursts are extremely sporadic or
not observed at all. Although WZ Sge-type stars were rare until recently, their number has increased dramatically since the
mid-2000s, with a discovery rate of a few dozen per year, approaching now to 300-400 objects.\footnote{Unfortunately, there
is currently no up-to-date catalogue of WZ~Sge-type DNe accessible to the community. To address this gap, we are in the process
of compiling a new catalogue tailored to the needs of our study (Neustroev et al., in prep). It already
includes more than 300 objects, more than triple the number since Kato's (2015) review \cite{kat15wzsge}.}
There is a strong reason to believe that this number ($\sim$10\% of a few thousand of currently known CVs) is just the tip
of an iceberg. According to standard evolutionary theory \cite{kni11CVdonor}, the majority of the present-day CVs should have
already evolved to short orbital periods and now concentrate close to the so-called period minimum \cite{bel20cvevol}. Being
faint and beyond our ability to detect or recognise them, they create a dormant population of accreting white dwarfs (AWDs).
However, such objects will eventually, sooner or later, undergo a superoutburst and become WZ Sge-type CVs if we are fortunate
enough to discover them.

Thus, WZ Sge-type stars are among the most populous subtypes of CVs. However, despite their great importance for a vast range
of astrophysical questions, the DIM still fail to reproduce their behaviour properly. To explain the large amplitude, long
recurrence time, and long duration of the superoutbursts, the DIM requires extremely low viscosity in the quiescent disc, with
$\alpha$$\lesssim$10$^{-4}$ \cite{sma93wzsge,osa95wzsge}, the physical reason for which is unknown. If one wants to keep
viscosity's value at a ``normal'' level of $\alpha$$\sim$0.01, then the inner disc should be sufficiently truncated and a
superoutburst must be triggered by enhanced mass transfer from the donor, which should persist during the superoutburst
\cite{ham97wzsgemodel,ham21V3101Cygrebrightening}.

In this conference paper, we probe the physical properties of accretion discs in a sample of WZ~Sge-type objects in deep
quiescence. We present some already published and also preliminary results of high-quality observations obtained with the
Very Large Telescope at the Paranal Observatory in Chile, using the medium-resolution spectrograph X-shooter. We complement
our observations with archival spectroscopic and photometric data obtained at other facilities such as the Nordic Optical
Telescope, the Hubble Space Telescope (HST), the Neil Gehrels Swift Observatory, using both the X-ray Telescope (XRT) and the
UV/Optical Telescope (UVOT), and Wide-field Infrared Survey Explorer (WISE). We show that observations are inconsistent with
certain predictions of the DIM.

\section{WZ Sge-type stars and ordinary SU UMa-type DNe in comparison}

Originally, the DIM was developed to reproduce the normal outbursts observed in ordinary DNe. Adopting the TTI model, it can also
satisfactorily explain superoutbursts in SU~UMa-type stars. However, as noticed above, the very long interval between
superoutbursts of a larger amplitude and a longer duration in WZ~Sge-type objects, and also rebrightenings (echo outbursts),
observed after the main superoutburst in many such objects \cite{kat15wzsge} (but not all!) require additional assumptions,
the physical reason for which is not known.

\begin{figure}[ht]
    \centering
    \includegraphics[width = 0.96\textwidth]{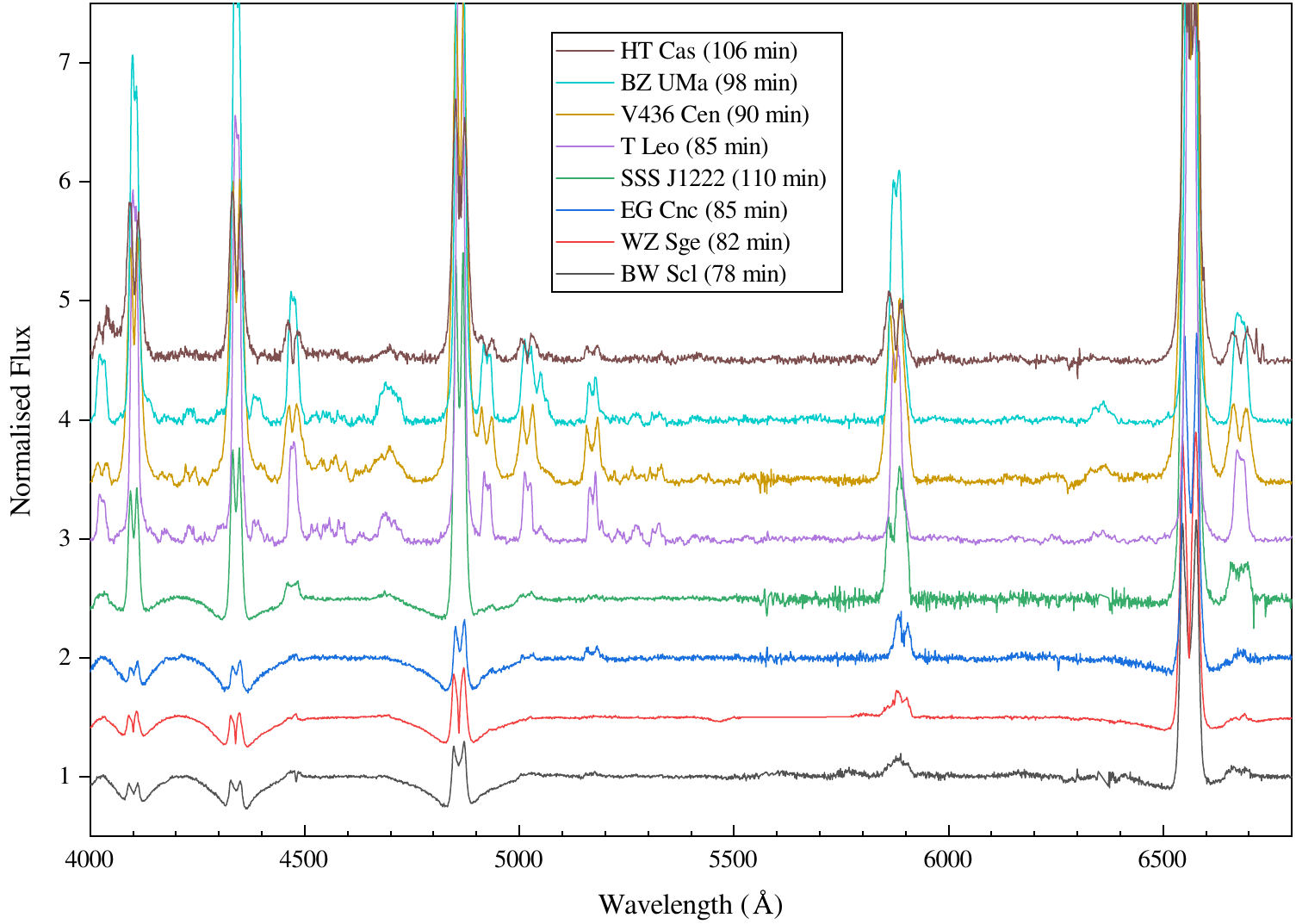}
    \caption{Normalised spectra of ordinary SU~UMa-type stars (HT~Cas, BZ~UMa, V436 Cen, and T~Leo -- four upper spectra) and
    WZ~Sge-type objects (SSS J122221.7-311525, EG~Cnc, WZ~Sge, and BW~Scl -- four lower spectra). The orbital periods of all
    the objects are indicated in the plot legend. The plot emphasises the Balmer absorption features and continuum, thus, the
    Balmer emission lines in some objects do not fit in the plot. The spectra of the WZ Sge-type objects clearly exhibit the
    presence of the broad Balmer absorption lines from a WD, whereas the ordinary SU~UMa-type stars do not.
    }
    \label{Fig:SU_UMa_spectra}
\end{figure}

What is the principal difference between accretion discs in these two types of DNe? Both types of binaries are essentially
the same: having similar orbital periods, they have a similar size, and their accretion discs must also be of a similar size.
Probably, the only significant difference is the mass-transfer rate $\dot{M}_{\rm tr}$ from the donor, at least during the
quiescent period. As the total accretion luminosity $L_{\rm acc}$ is directly proportional to $\dot{M}_{\rm tr}$, the accretion
discs in lower $\dot{M}_{\rm tr}$ systems must be less luminous. This is what we indeed observe in optical spectra of DNe.

\begin{figure}
    \centering
    \includegraphics[width = 0.49\textwidth]{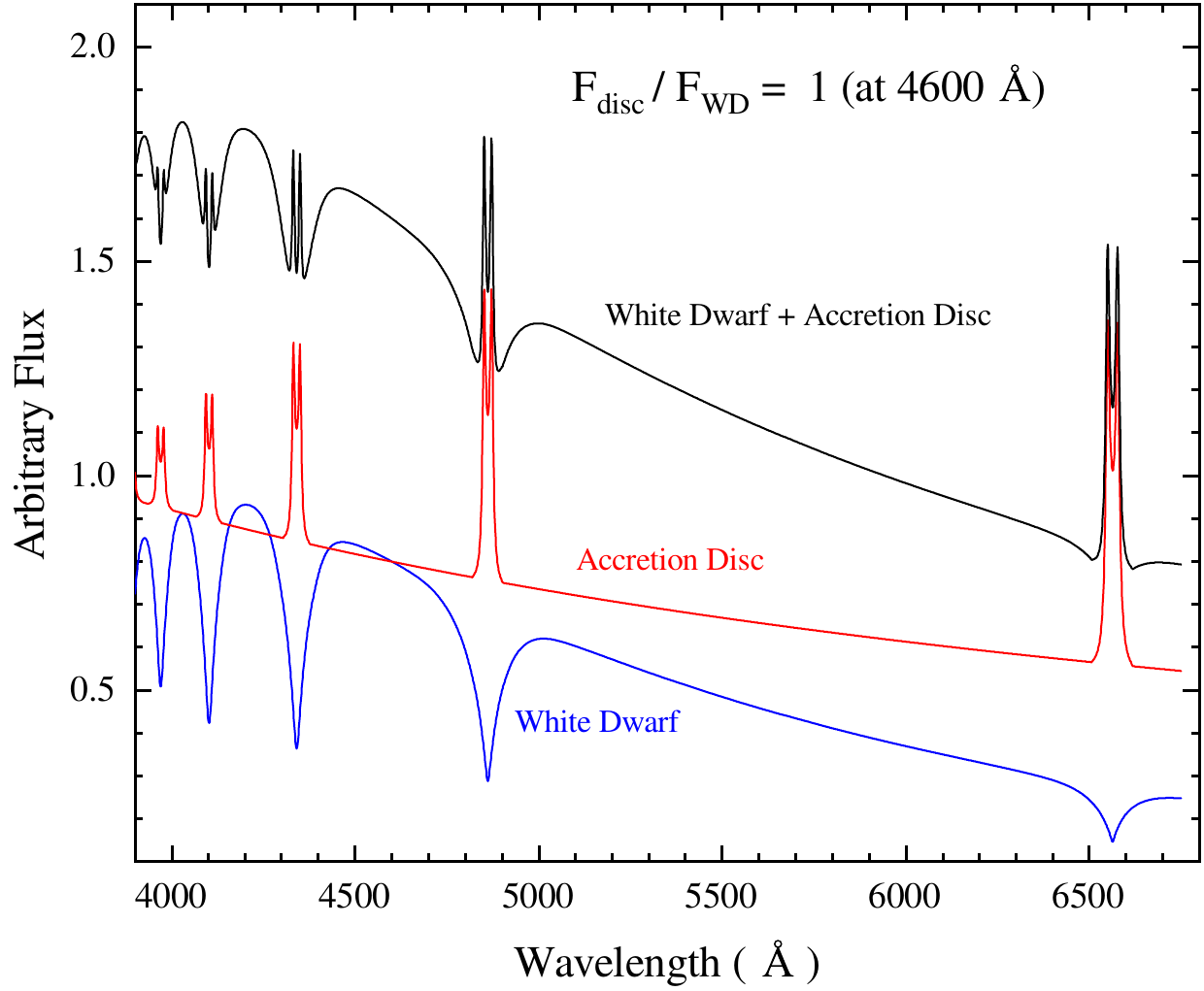}
    \includegraphics[width = 0.49\textwidth]{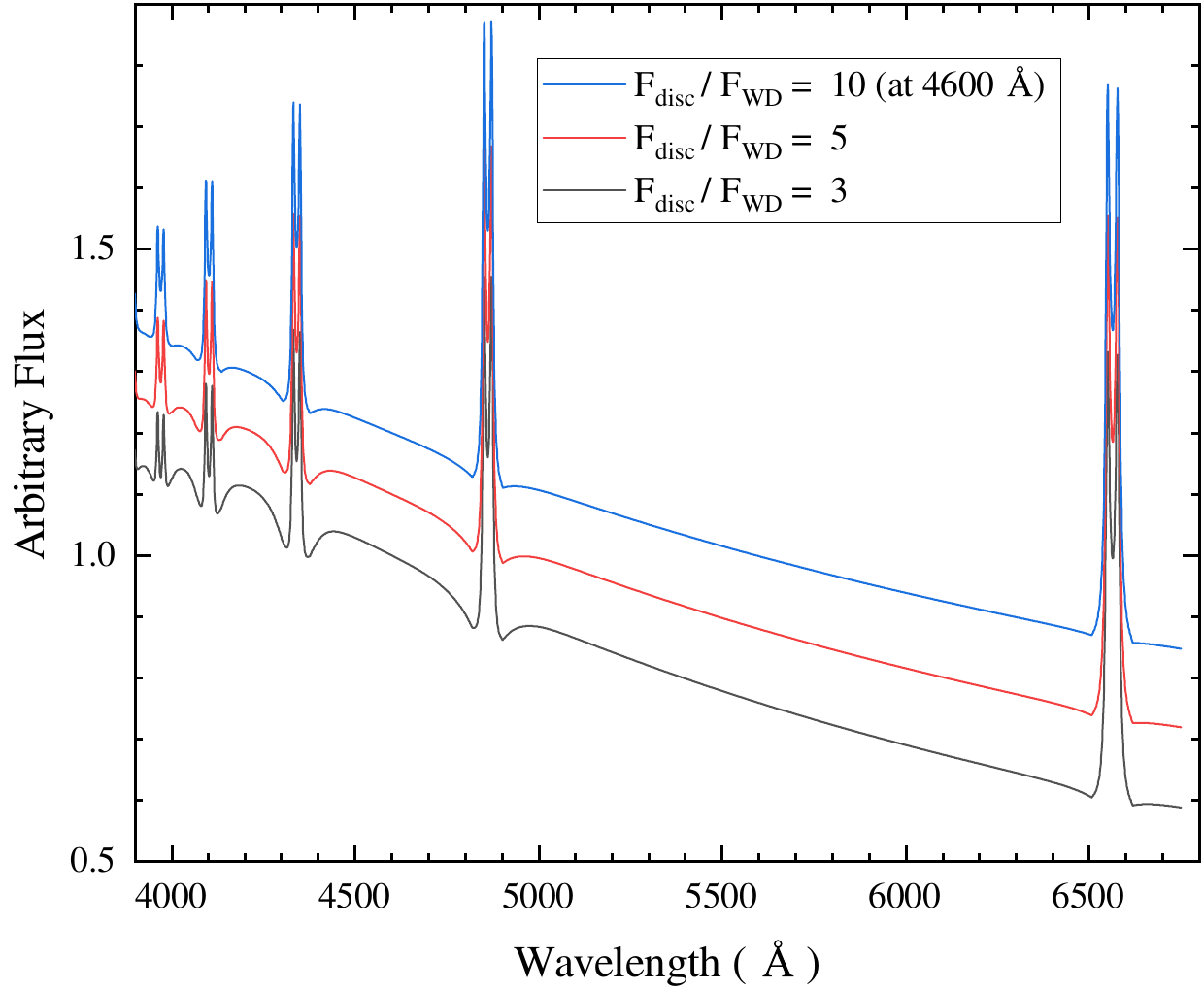}
    \caption{Simulated spectra of a CV calculated as a sum of a WD model spectrum and a power law with added double-peaked
    emission lines of arbitrary strength, mimicking an accretion disc spectrum. The left panel shows the WD and disc spectra
    with the same flux at 4600 \AA, along with their combination. The right panel shows three combined spectra with different
    flux ratios.}
    \label{Fig:ADandWD}
\end{figure}

Figure~\ref{Fig:SU_UMa_spectra} shows normalised spectra of four ordinary SU~UMa-type stars (HT~Cas, BZ~UMa, V436 Cen, and T~Leo)
and four WZ~Sge-type objects (SSS J122221.7-311525, hereafter SSS J1222, EG~Cnc, WZ~Sge, and BW~Scl). Both samples cover similar
ranges of orbital
periods. We note that the spectra of the ordinary SU~UMa-type stars are similar in appearance to those of longer-period DNe, such
as SS~Cyg or U~Gem. They are dominated by hydrogen and neutral helium emission lines superposed on a smooth continuum. In contrast,
the WZ~Sge-type spectra exhibit much fainter helium lines, and, more importantly, the clear presence of broad Balmer absorption
lines from the WD. The latter is a direct indication of much lower disc luminosity in the WZ~Sge-type objects than in other classes
of DNe. It is easy to verify that in order to hide the underlying WD absorption lines, the accretion disc must be at least 10 times
brighter than the WD (Figure~\ref{Fig:ADandWD}), while various studies indicate that in the WZ~Sge-type objects the disc contributes
only a few dozen per cent of the total system light in the blue optical spectrum \cite{neu23bwscl}.

\begin{figure}
    \centering
    \includegraphics[width = 0.49\textwidth]{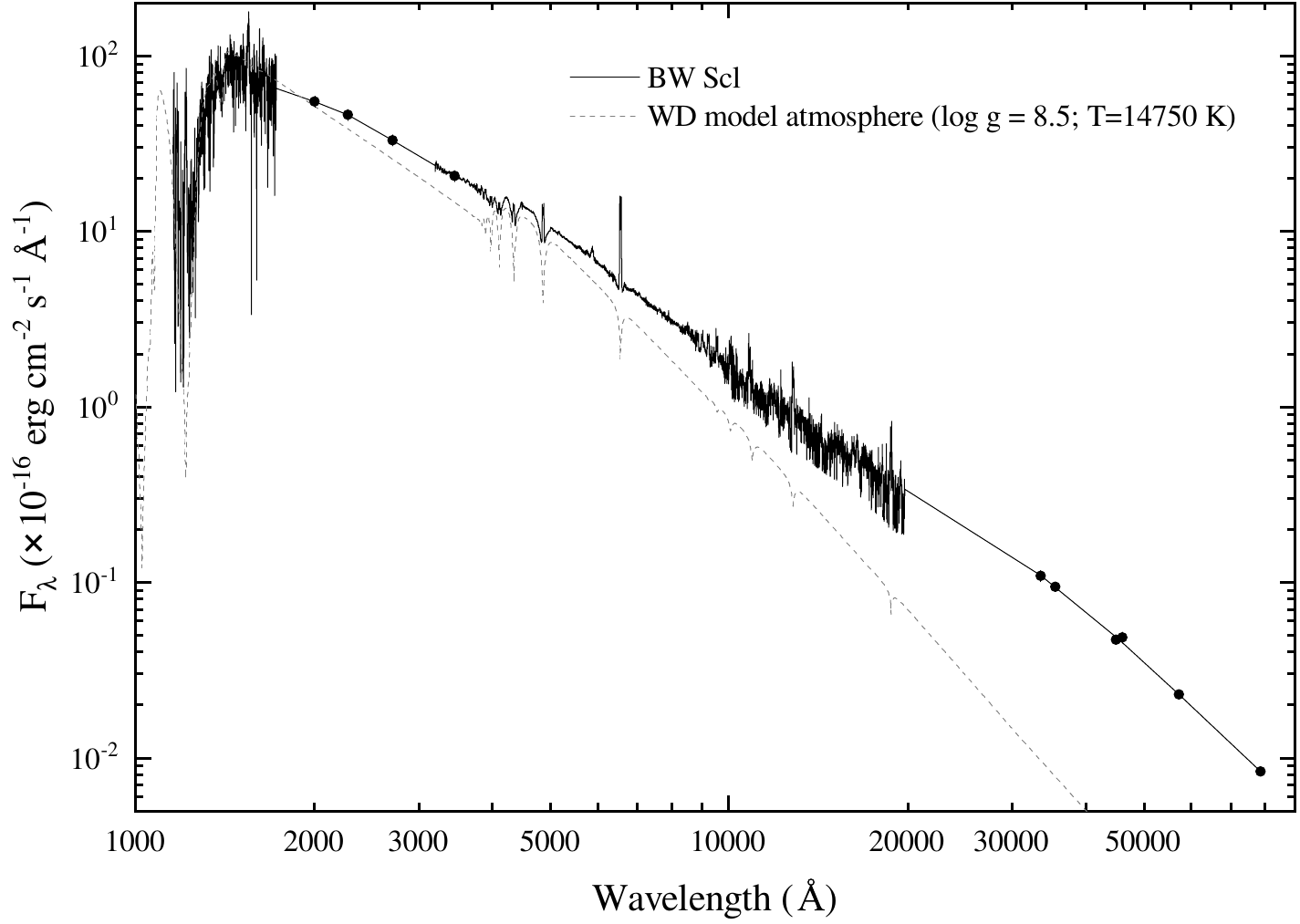}
    \includegraphics[width = 0.49\textwidth]{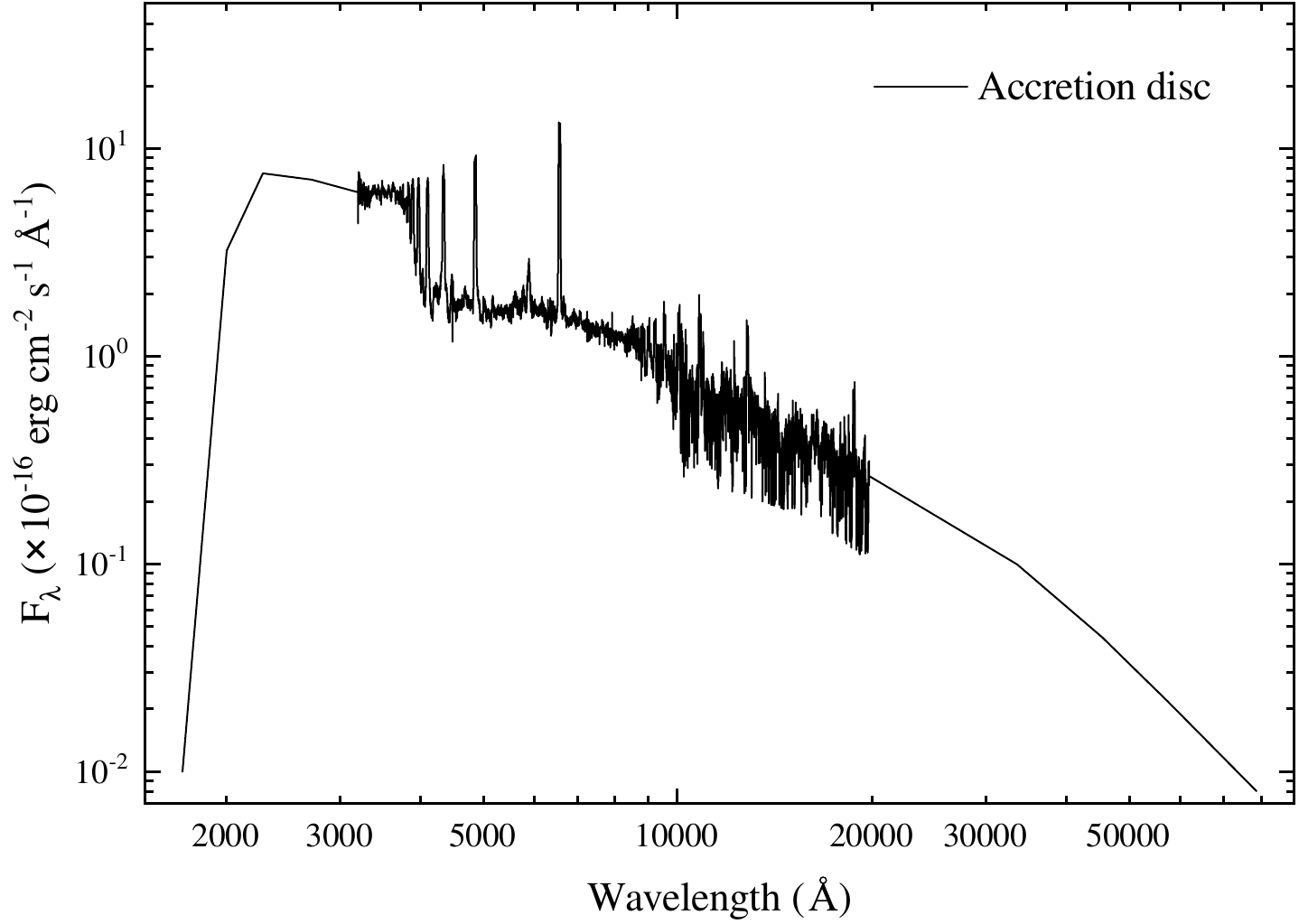}
    \caption{Left: the spectrum of BW Scl, shown together with the best-fitting WD model spectrum.
    The WD emission (grey dashed line) contributes 90 cent of the total system flux at 4600 \AA.
    Right: The accretion disc spectrum obtained after a subtraction of the WD spectrum from the spectrum of BW Scl.
    Its colour temperature is 10580$\pm$40~K.}
    \label{Fig:SED}
\end{figure}

\section{Accretion disc spectrum}

Thus, accretion discs in WZ Sge-type stars have a very low luminosity.  What physical characteristics do these disks have? Even
the spectra (continuum) of such discs are poorly known, making it challenging to determine their properties. The optical spectrum
is hard to derive because the WD dominates the system spectrum.
To extract the accretion disc spectrum, we used the approach described in \cite{neu23bwscl}, which provides all details. Shortly,
we first derive the WD parameters ($\log g$ \& $T_{\rm eff}$) by fitting the object spectrum between the \Hbeta\ and \Hdelta\
lines to a grid of synthetic spectra of DA WDs, to which the power-law or blackbody flux (mimicking a disc contribution) was
added. In this range of wavelengths, the contribution of the disc is minimal (Figure~\ref{Fig:SED}, left), enabling the fitting
procedure to constrain both the WD parameters and the WD contribution to the total light \cite{JussiThesis}.
We then subtract the found underlying WD spectrum from the object’s broadband spectrum. The latter is constructed from
a combination of the optical spectrum, an archival HST UV spectrum and/or Swift/UVOT photometry in the UV, and WISE photometry
in the mid-NIR. In most cases, we used an X-shooter spectrum spanning a broad range of wavelengths from the near-UV to the
near-infrared. The resulting spectrum can be considered the accretion disc spectrum. Two examples of the extracted disc spectra
of BW Scl and EG~Cnc can be seen in Figures~\ref{Fig:SED} (right) and ~\ref{Fig:EG_Cnc_Aver}.

We must note, however, that the obtained spectrum is a combination of contributions from the disc itself, the hotspot, and
the donor star. The latter mostly contributes in the NIR, and, for all the considered objects here, this contribution is
negligible, as the donor is a brown-dwarf-like object (e.g., \cite{neu17j1222,neu23bwscl}). On the other hand, the hotspot
can be a major contributor, having luminosity even higher than that of the disc itself. In the following calculations, we will
assume that their contributions are equal, as we are currently unable to separate the disc and hotspot shares.

\section{Probing accretion disc physical parameters}
\subsection{Bolometric Luminosity, Mass-Transfer, and Mass-Accretion Rates}

Using the above approach, we extracted the disc spectra from 5 WZ~Sge-type stars. For BW~Scl and EG~Cnc, two spectra were obtained
using observations performed before and after their most recent superoutbursts (in 2010 and 2018, respectively). By integrating the
disc SEDs over all wavelengths and adopting the \textit{Gaia} distances, we can put a conservative upper limit on the bolometric
luminosity of the disc $L_{\rm d, bol}$ (Table~\ref{Tab:ADsed}). All the objects were also observed with the Swift/XRT, allowing
us to estimate the unabsorbed X-ray luminosity $L_{\rm x}$, also showed in the table.\footnote{Assuming that X-ray spectra of
these objects are similar to other WZ Sge-type stars such as GW~Lib and SSS~J1222 \cite{neu18j1222gwlib}, we estimated $L_{\rm x}$
in the 0.3–10 keV range by using the count-rates as the scale-factor.} Note that in most cases, $L_{\rm x}$ < $L_{\rm d, bol}$.

The found values of $L_{\rm d, bol}$ can be used to evaluate the mass-accretion $\dot M_{\rm acc}$ and mass-transfer rate $\dot M_{\rm tr}$.
We calculated $\dot{M}_{\rm acc}$ using
\begin{equation}
 \dot M_{\rm acc} = \frac{2 \xi L_{\rm d, bol} R_{\rm wd}}{G M_{\rm wd}} \,,
\label{Eqn:Macc}
\end{equation}
where $R_{\rm wd}$ and $M_{\rm wd}$ are the radius and mass of the WD, G is the gravitational constant, and $\xi$=0.5 is
the adopted fraction of the system luminosity produced by the accretion disc. Note that Eqn~(\ref{Eqn:Macc}) is valid for
steady accretion, while for quiescent discs it provides an upper limit on $\dot M_{\rm acc}$.

$\dot{M}_{\rm tr}$ can be estimated from the luminosity of the hotspot, which is $(1-\xi)L_{\rm d, bol}$. The luminosity is
determined by the release of gravitational potential energy as material falls from infinity to the impact point at the rim of
the accretion disc \cite{Warner}, or by the conversion of the gas stream's kinetic energy into radiation
\cite{sma02ADstructure,sma05ugemEMT}; both are proportional to $\dot{M}_{\rm tr}$. Then we can use one of the following equations:
\begin{equation}
 \dot M_{\rm tr} \approx \frac{(1-\xi)}{\eta}\frac{r_{\rm hs}L_{\rm d, bol}}{G M_{\rm wd}} \approx \frac{2 (1-\xi)}{\eta}\frac{L_{\rm d, bol}}{\Delta V^2}
 \,,
\label{Egn:Mtr}
\end{equation}
where $r_{\rm hs}$ is a disc radius at which the hotspot is located, $\Delta V$ is the impact velocity of the gas stream relative
to the disc (then $\Delta V^2$/2 is the energy dissipation per 1 gram of the stream material), and $\eta$ is an efficiency factor.
With $\eta$<1, one can take into account that (in the first equation) the material falls from the inner Lagrangian point, not
from $\infty$, that the stream meets the rotating disc obliquely, and that (in the second equation) not all kinetic energy is
converted into the luminosity of the spot. In order to be on the conservative side, we adopt $\eta$=0.5 below.

Using the second equation and assuming $\eta$=1, Smak \cite{sma93wzsge} estimated $\dot{M}_{\rm tr}$ in WZ~Sge to be 2\tim{15}
g s$^{-1}$. This value remains, in fact, the only measurement of this parameter in this type of DNe and is now commonly adopted
for the whole class of WZ~Sge-type stars, including within the DIM \cite{bua02suumamodel}. However, Smak's calculations were
based on parameters later shown to be incorrect, leading to a notable overestimation of the result. First and foremost, the WD
mass in WZ~Sge is actually about twice as large, which affects the predicted impact velocity of the gas stream. $\Delta V$ was
determined from the kinematics of the stream colliding with the outer rim of the disc. The latter was evaluated from a
peak-to-peak separation of double-peaked emission line profiles. However, we show below that this approach is incorrect, as the
hotspot is located well inside the disc. Finally, our hotspot luminosity is at least 2 times lower than that of Smak, who, based
on a number of assumptions and adopting a bolometric correction, arrived at a higher value.

We must note that the structure of the hotspot in WZ~Sge-type stars is very complex. As we show in Section~\ref{Sec:Dopmaps}
(for more detail see \cite{neu23bwscl}), the hotspot is highly elongated. Although it becomes visible at the disc edge, its
brightest part
is located well inside the disc, close to its circularisation radius. It makes the use of any of the Equations (\ref{Egn:Mtr})
difficult as both the most important parameters $r_{\rm hs}$ and $\Delta V$ appear to be ambiguous. We believe that taking
into account the reduction in stream speed while the stream is moving through the low-density disc is less certain. For this
reason, we use the first formula with $r_{\rm hs}$ having a value of a brightness-weighted centroid of the hotspot estimated
from the Dynamical Doppler maps by eye (Section~\ref{Sec:Dopmaps}). In most cases $r_{\rm hs}$ is close to 0.40$a$, and only
for EG~Cnc we used 0.45$a$, where $a$ is the binary separation.

Table~\ref{Tab:ADsed} shows our results. We find that all the obtained values are very consistent, being roughly
$\dot{M}_{\rm acc}$ $\lesssim$ (1$\div$2)\tim{13} g s$^{-1}$ $\approx$ (2$\div$3)\tim{-13} \Msun\ yr$^{-1}$ and
$\dot{M}_{\rm tr}$ $\approx$ (2$\div$4)\tim{14} g s$^{-1}$ $\approx$ (3$\div$6)\tim{-12} \Msun~yr$^{-1}$,
about a factor of 10 smaller than those commonly assumed for WZ~Sge-type stars.

\begin{table}
\begin{center}
\caption{Bolometric and unabsorbed X-ray luminosities (in the 0.3–10 keV range) of the disc and the mass-accretion and
mass-transfer rates in a sample of WZ~Sge-type DNe.}
\begin{tabular}{lccccc}
\hline \hline
Object    & Year of     & $L_{\rm d, bol}$ & $L_{\rm x}$    & $\dot{M}_{\rm acc}$           &  $\dot{M}_{\rm tr}$ \\
          & observation & \tim{30} \erg    & \tim{30} \erg & \tim{13} g s$^{-1}$ &  \tim{13} g s$^{-1}$ \\
\hline
BW Scl    & 2010 & 3.2 &      &  $\lesssim$2.0 &    45.8 \\
          & 2017 & 4.0 & 0.86 &  $\lesssim$2.5 &    57.3 \\
EG Cnc    & 2003 & 1.3 & 0.61 &  $\lesssim$0.6 &    20.1 \\
          & 2019 & 2.2 &      &  $\lesssim$0.9 &    34.0 \\
EZ Lyn    & 2018 & 1.5 & 0.14 &  $\lesssim$0.9 &    23.0 \\
SSS J1222 & 2019 & 2.0 & 3.09 &  $\lesssim$1.2  &   36.0 \\
WZ Sge    & 2025 & 1.6 & 0.94 &  $\lesssim$1.0 &    23.9 \\
\hline
\end{tabular}
\label{Tab:ADsed}
\end{center}
\end{table}

\subsection{Physical properties of the accretion discs}

A drastic difference in disc luminosities in SU~UMa-type and WZ~Sge-type stars indicates that the discs in the latter might be
in an optically thin regime. Indeed, it has been shown \cite{wil80ADemissionline} that accretion discs in CVs with low mass
accretion rates should have outer regions optically thin in continuum, and that at $\dot{M}_{\rm acc}$ = 5$\times$$10^{13}$
g s$^{-1}$ the entire disc becomes optically thin in continuum \cite{Tylenda81}. We found above that the observed
$\dot{M}_{\rm acc}$ in the studied objects is lower than this limit (Table~\ref{Tab:ADsed}).

As a simple exercise, we can estimate the mean effective (blackbody) temperature $T_{\rm eff}$ of the disc using the definition
of the luminosity $L_{\rm d}$ as the integral of the total flux over the disc surface:

\begin{equation}
 L_{\rm d} = 2 \pi (r_{\rm out}^2-r_{\rm in}^2) \sigma T_{\rm eff}^4 \,,
\end{equation}
where $r_{\rm out}$ and $r_{\rm in}$ are the outer and inner radii of the disc and $\sigma$ is the Stefan-Boltzmann constant.
Assuming that $L_{\rm d}$=0.5$L_{\rm d, bol}$ (another half is emitted by the hotspot) and using the disc parameters found for
BW~Scl \cite{neu23bwscl} ($L_{\rm d}$=1.6\tim{30} \erg, $r_{\rm in}$=0.01 \Rsun$\approx$$R_{\rm wd}$, and $r_{\rm out}$=0.338
\Rsun), we obtain $T_{\rm eff}$=1570 K (similar and even lower values are obtained for other objects). It is unlikely that such
a low $T_{\rm eff}$ represents the true, kinetic temperature of the disc material, as no strong Balmer emission lines can be
produced at such a low temperature, and the spectrum should be peaked at NIR wavelengths, thus providing additional support for
the optically thin conditions in the disc.

Can a part of the disc still be in the optically thick regime? It is easy to show that even if it is so, such optically thick
rings must be unrealistically narrow.  Assuming the above parameters and that a ring with even relatively low
$T_{\rm eff}$=4000~K produces the whole disc luminosity $L_{\rm d}$ (i.e. the optically thin part emits nothing), its size will
be (1$\div$6)~$R_{\rm wd}$, if it is located at the inner edge, or (0.984$\div$1)~$R_{\rm out}$ = (33.3$\div$33.8)~$R_{\rm wd}$,
if it is at the outer disc edge. It should be emphasised that a blackbody spectrum with $T_{\rm eff}$=4000~K is already too red
to match the observed disc SED colour temperatures, which consistently fall within a relatively narrow interval of 8500–10500~K
across various objects and data sets (see, e.g., Figure~\ref{Fig:SED}, right). Furthermore, the strong Balmer jump and emission
lines observed in the spectrum indicate that the second assumption is also overly simplistic, further narrowing the optically
thick ring. Thus, based on the above arguments, we conclude that most of the accretion discs in WZ~Sge-type objects, possibly
the entire discs, are optically thin.

\begin{table}
\begin{center}
\caption{The EWs of the Balmer emission lines and the Balmer decrement (BD), measured in the accretion disc spectra of a sample
of WZ~Sge-type systems.}
\begin{tabular}{lcrrrrc}
\hline \hline
Object  & Year of        &\multicolumn{4}{c}{Equivalent Width (\AA)}& Balmer Decrement \\
Object  & observation     &\Halpha&\Hbeta&\Hgamma&\Hdelta&  \Halpha\ : \Hbeta\ : \Hgamma\ : \Hdelta\\
\hline
BW Scl                   & 2010  & 341 & 210 & 190 & 170 & 1.71 : 1.00 : 0.86 : 0.71 \\
--- $\shortparallel$ --- & 2017  & 340 & 194 & 178 & 174 & 1.60 : 1.00 : 0.89 : 0.82 \\
EG Cnc (with hotspot)    & 2019  & 280 & 121 &  97 & 140 & 2.31 : 1.00 : 0.63 : 0.53 \\
--- $\shortparallel$ --- (without hotspot) & & 265 &  96 &  79 &  74 & 2.72 : 1.00 : 0.60 : 0.48 \\
EZ Lyn                   & 2018  & 703 & 205 & 141 & 102 & 1.80 : 1.00 : 0.71 : 0.49 \\
SSS J1222                & 2019  & 842 & 332 & 211 & 164 & 2.10 : 1.00 : 0.63 : 0.45 \\
WZ Sge                   & 2025  & 303 & 118 &  79 &  73 & 1.86 : 1.00 : 0.74 : 0.65 \\
\hline
\end{tabular}
\label{Tab:EW}
\end{center}
\end{table}

What are the physical conditions within such accretion discs? Unfortunately, there is no established methodology capable
of directly constraining them in such discs. However, some clues about the temperature and density of the line-emitting
regions can be evaluated from emission lines and the continuum shape. Having
recovered not only a non-WD continuum but also higher-order Balmer emission lines that lie within the WD absorption troughs,
we can use their equivalent widths (EWs) and the Balmer decrement to infer the temperature and density. It has been shown
that the Balmer decrements (especially \Halpha/\Hbeta) are sensitive to a great extent to a gas density in the optically thin
regime \cite{WilliamsShipman88}. Comparing the measured parameters of the Balmer lines (Table~\ref{Tab:EW}) with the model
predictions calculated by Williams \cite{Williams91}, we find that they are roughly consistent with the disc temperature in
the range of 8\,000--10\,000 K and the number density of hydrogen at the midplane of $\log N_0 \approx 12$ that corresponds
to the surface density $\Sigma$ of 0.006 g cm$^{-2}$.

\begin{figure}
    \centering
    \includegraphics[width = 0.49\textwidth]{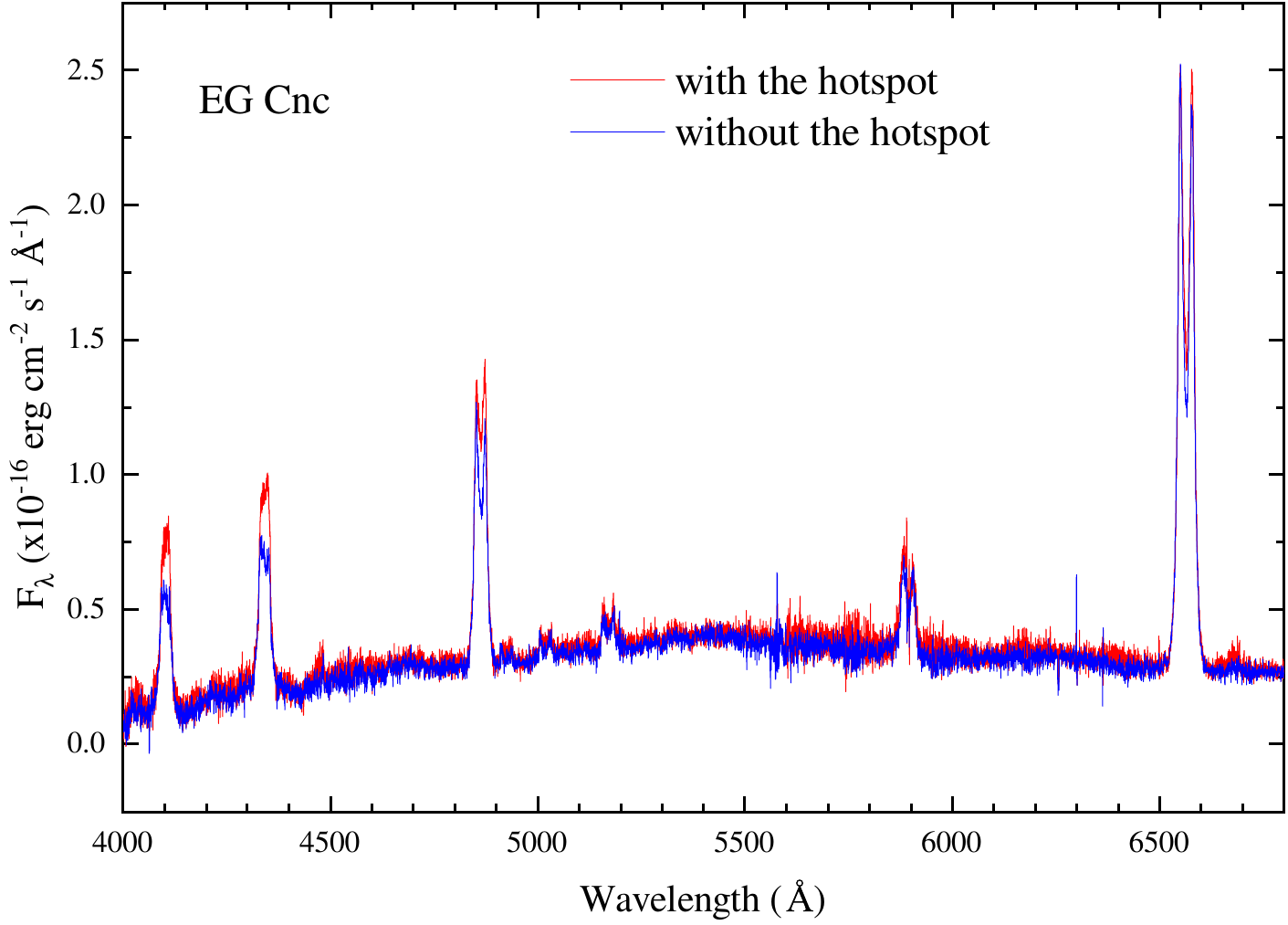}
    \includegraphics[width = 0.49\textwidth]{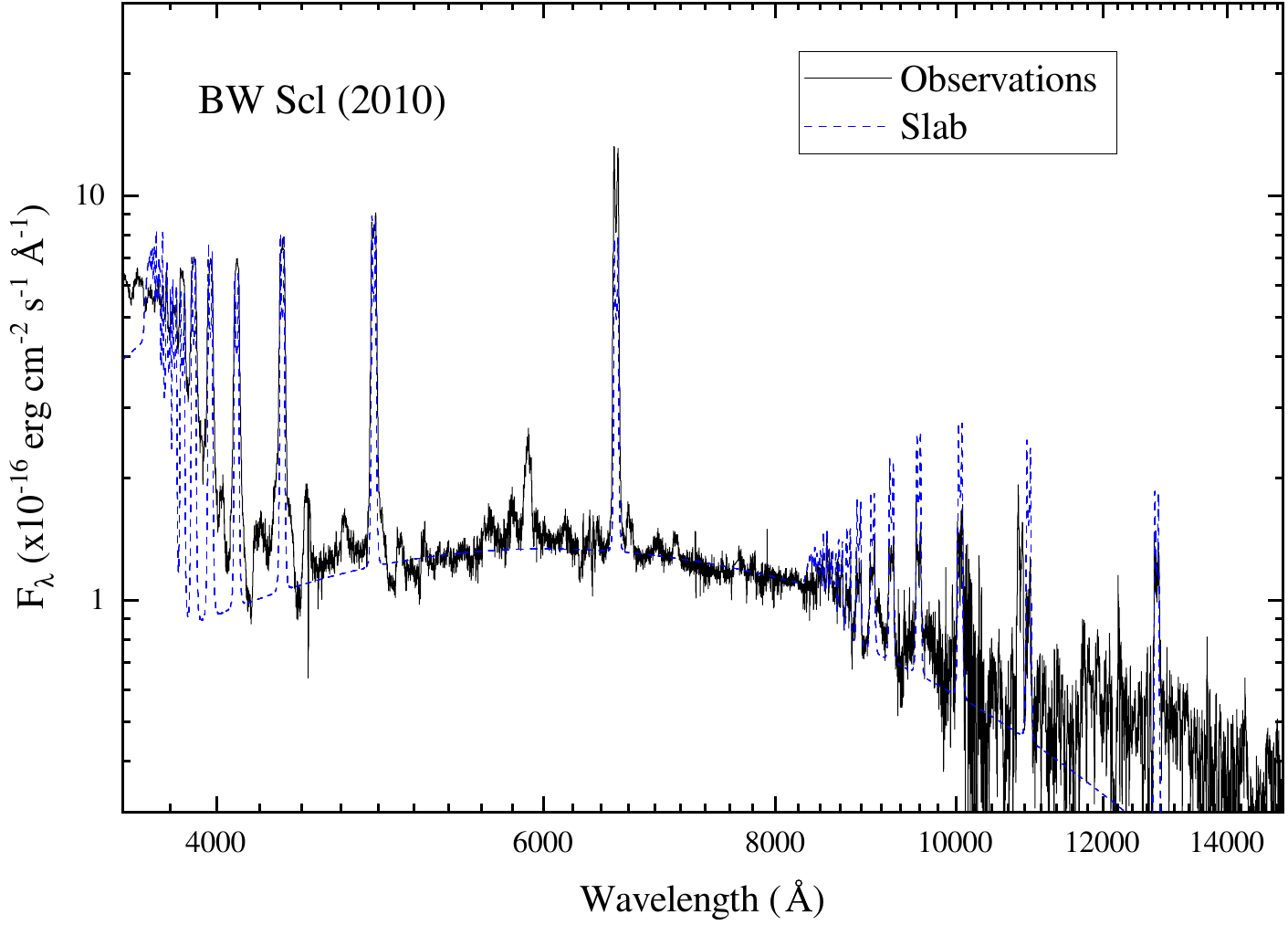}
    \caption{Left: Accretion disc spectra of EG Cnc with (red) and without (blue) hotspot contribution to the Balmer lines.
    Right: Accretion disc spectrum of BW Scl (black solid line) and an example (not the best-fit) of a hydrogen slab model
    (blue dashed line) calculated with the code of V. Suleimanov.
    }
    \label{Fig:EG_Cnc_Aver}
\end{figure}

It is important to note that the line parameters from Table~\ref{Tab:EW} are heavily affected by the presence of the
hotspot component in the lines. As we show below in Section~\ref{Sec:Dopmaps}, the hotspot feature seen in trailed spectra as
an S-wave becomes progressively stronger in higher-order Balmer lines in comparison with the underlying double-peaked line
originating in the accretion disc. This makes the measured Balmer decrement flatter than that of the disc spectrum alone.
It is not straightforward to properly separate the hotspot's contribution to the disc continuum, but it is relatively easy for
spectral lines, using the following approach. In Section~\ref{Sec:Dopmaps}, we show that, although the hotspot has a complex,
elongated structure, it remains relatively compact. Using time-resolved observations, we calculate two spectra, averaged
between orbital phases 0.4 and 0.8, when the hotspot S-wave is blue-shifted, and between phases 0.9 and 1.3, when it is
red-shifted. We then combine the right, red half of the line profiles from the first averaged spectrum with the left, blue half
from the second. We applied this method to the data for EG~Cnc and show the result in Figure~\ref{Fig:EG_Cnc_Aver} (left),
and the values of the calculated EW and the Balmer decrement in Table~\ref{Tab:EW}.
After removing the hotspot contribution to the emission lines, the Balmer decrement in the resultant disc spectrum appears
steeper, indicating even lower hydrogen density ($\log N_0 \approx 11.5-12$ and $T$$\approx8\,000-10\,000$ K) \cite{Williams91}.

Another method to assess the physical properties of accretion discs is to fit the whole disc spectrum with a hydrogen slab
\cite{her19V1838Aql} (see the right panel of Figure~\ref{Fig:EG_Cnc_Aver} for an example).
We used this approach to probe the disc in WZ Sge.  We found that the WD-subtracted spectrum must be fitted with separately
modelled slabs for the hotspot and disc contributions. We found that the disc has an even lower temperature of 5400~K and
a surface density of 0.03 g cm$^{-2}$. More details can be found in \cite{KvistDIM50} (these proceedings).

We should note that the obtained values of $\Sigma$ are likely underestimated, since with the measured $\dot{M}_{\rm tr}$,
such surface densities would be reached within a relatively short time. One possible explanation is that both methods rely
primarily on the properties of emission lines and therefore probe only the line-emitting regions, leaving the continuum-dominated
emitting regions unaccounted for.

\subsection{Double-peaked emission line profiles}

Some important accretion disc parameters can be constrained from modelling of double-peaked emission line profiles, which are
commonly observed in DNe and other types of CVs. A peak-to-peak separation in these profiles is
defined by the velocity of the outer rim of the accretion disc $V{\rm _{out}}$, while the ratio of the disc inner and outer radii
$r_{\rm in}/r_{\rm out}$ defines a full width of the line profile at zero intensity (FWZI) \citep{Smak69}. If the WD mass and
a system inclination are known, the above parameters can be converted to the outer and inner disc radii, or vice versa.

\begin{figure}
    \centering
    \includegraphics[width = 0.45\textwidth]{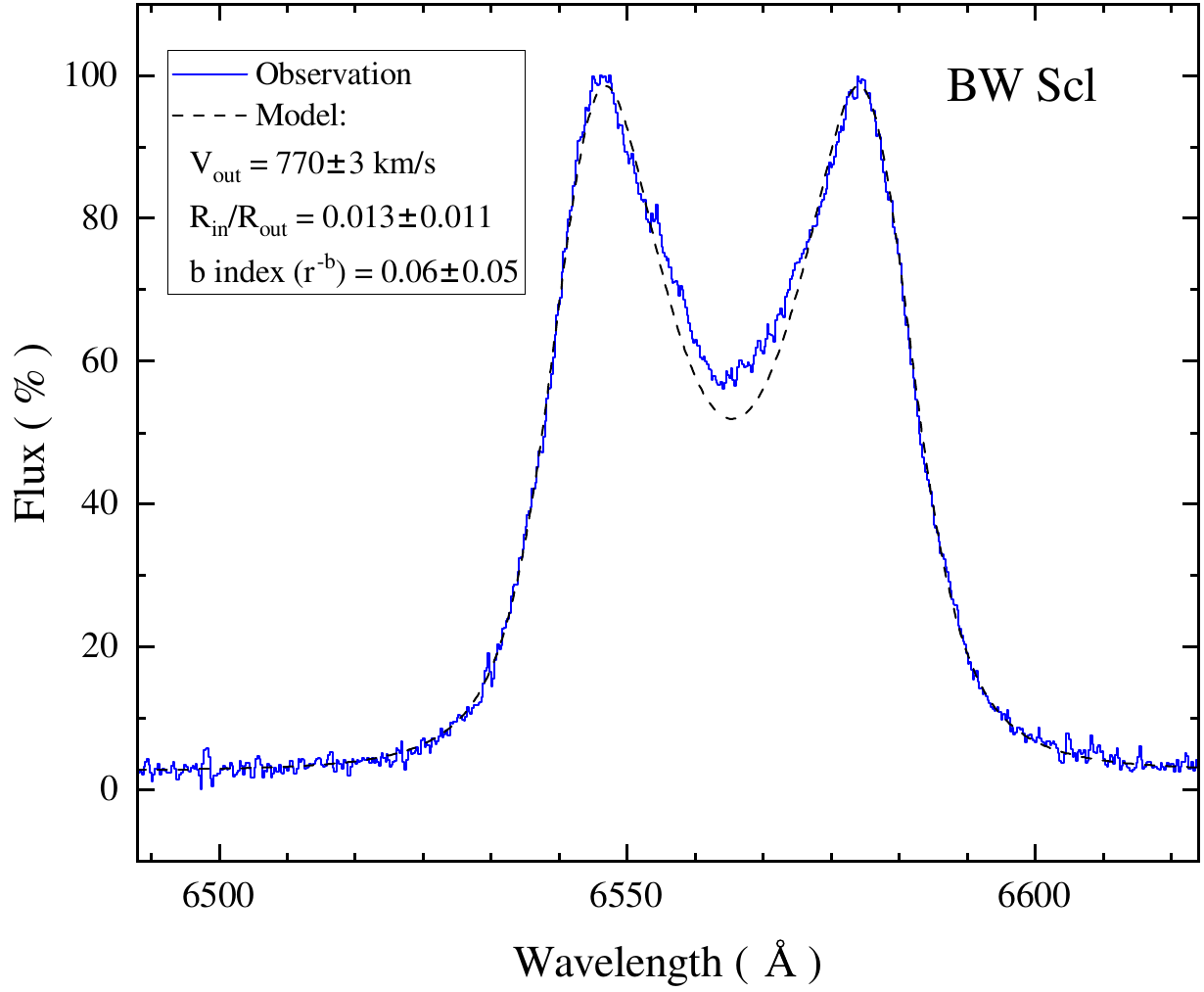}
    \includegraphics[width = 0.45\textwidth]{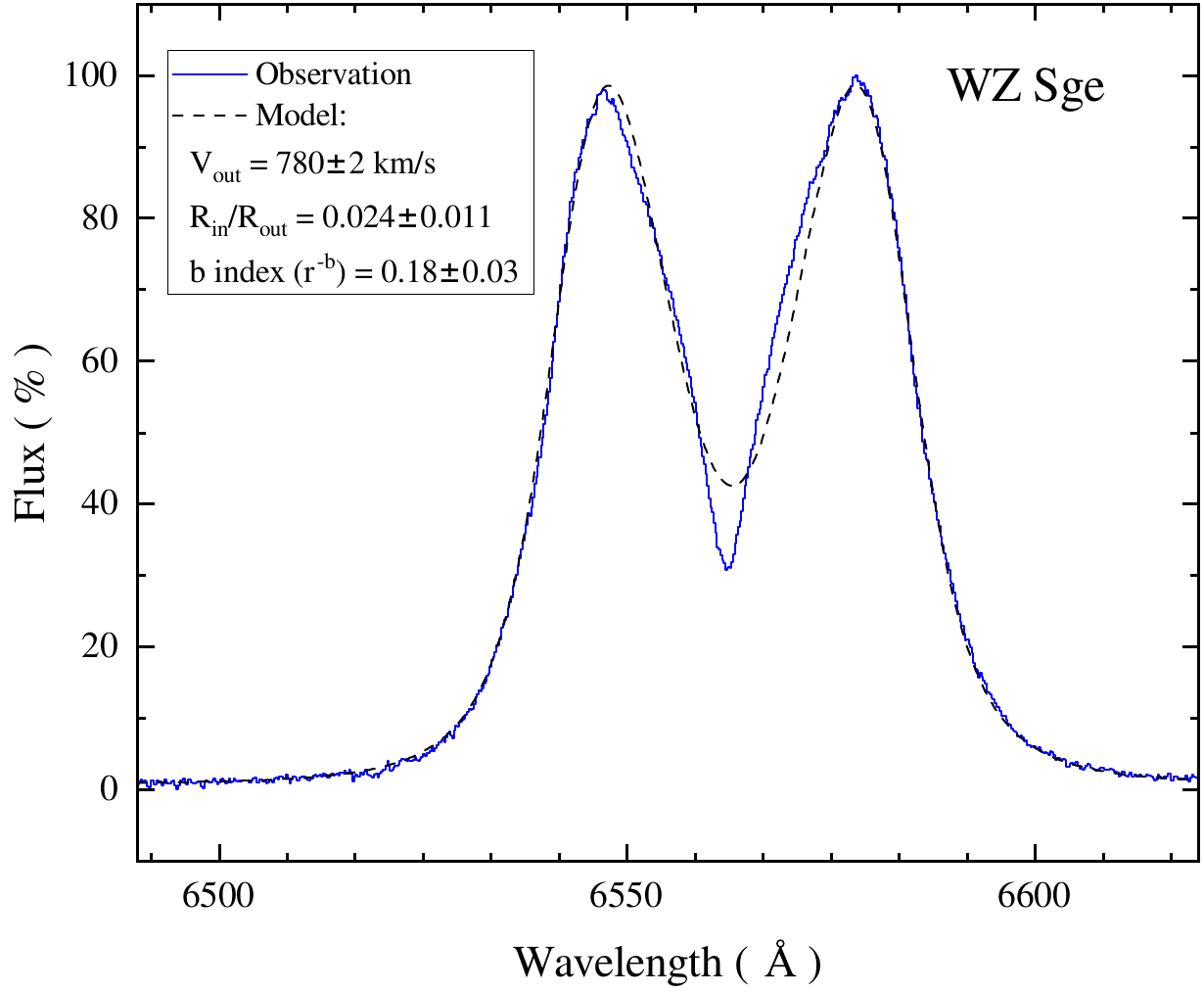} \\
    \includegraphics[width = 0.45\textwidth]{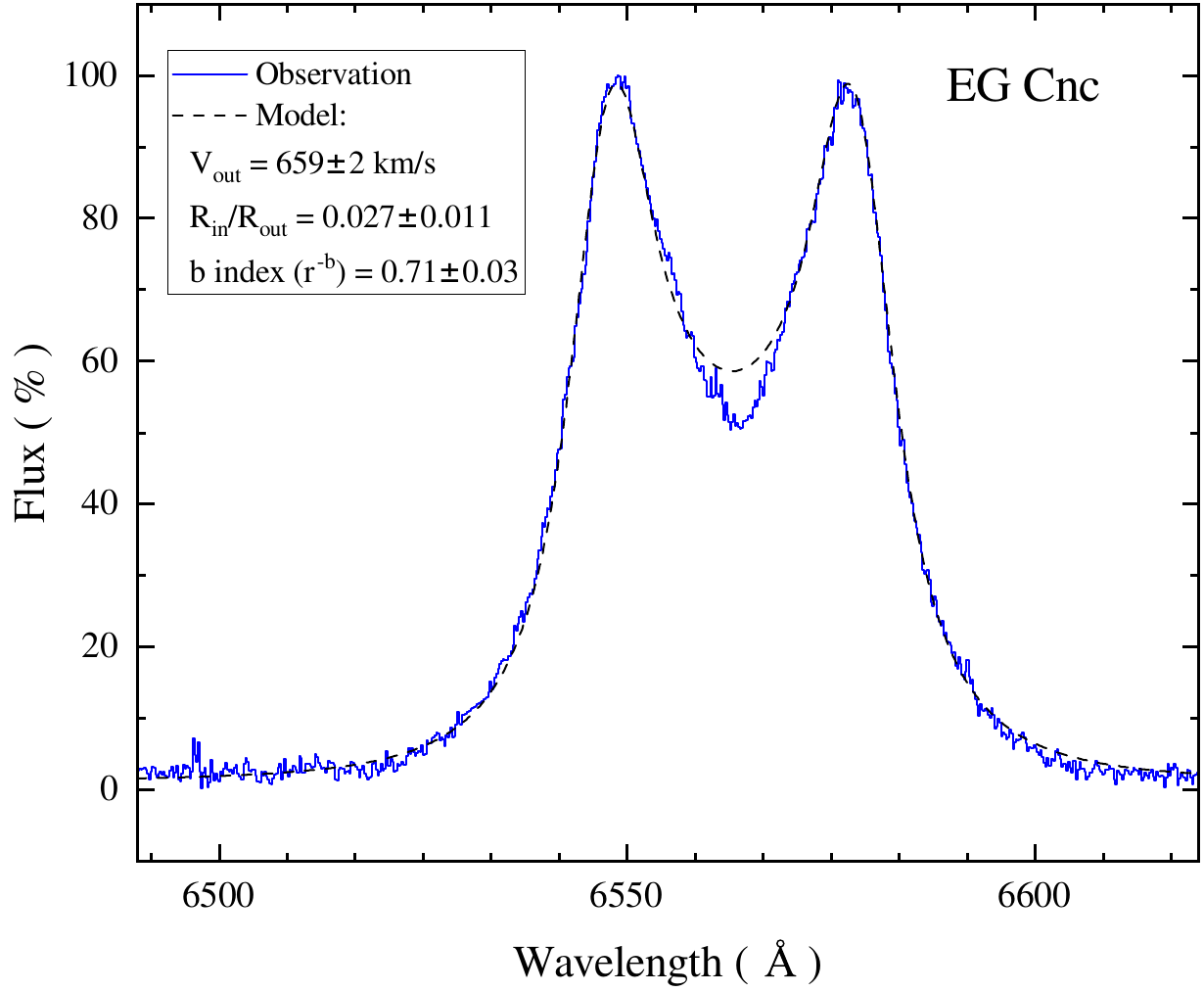}
    \includegraphics[width = 0.45\textwidth]{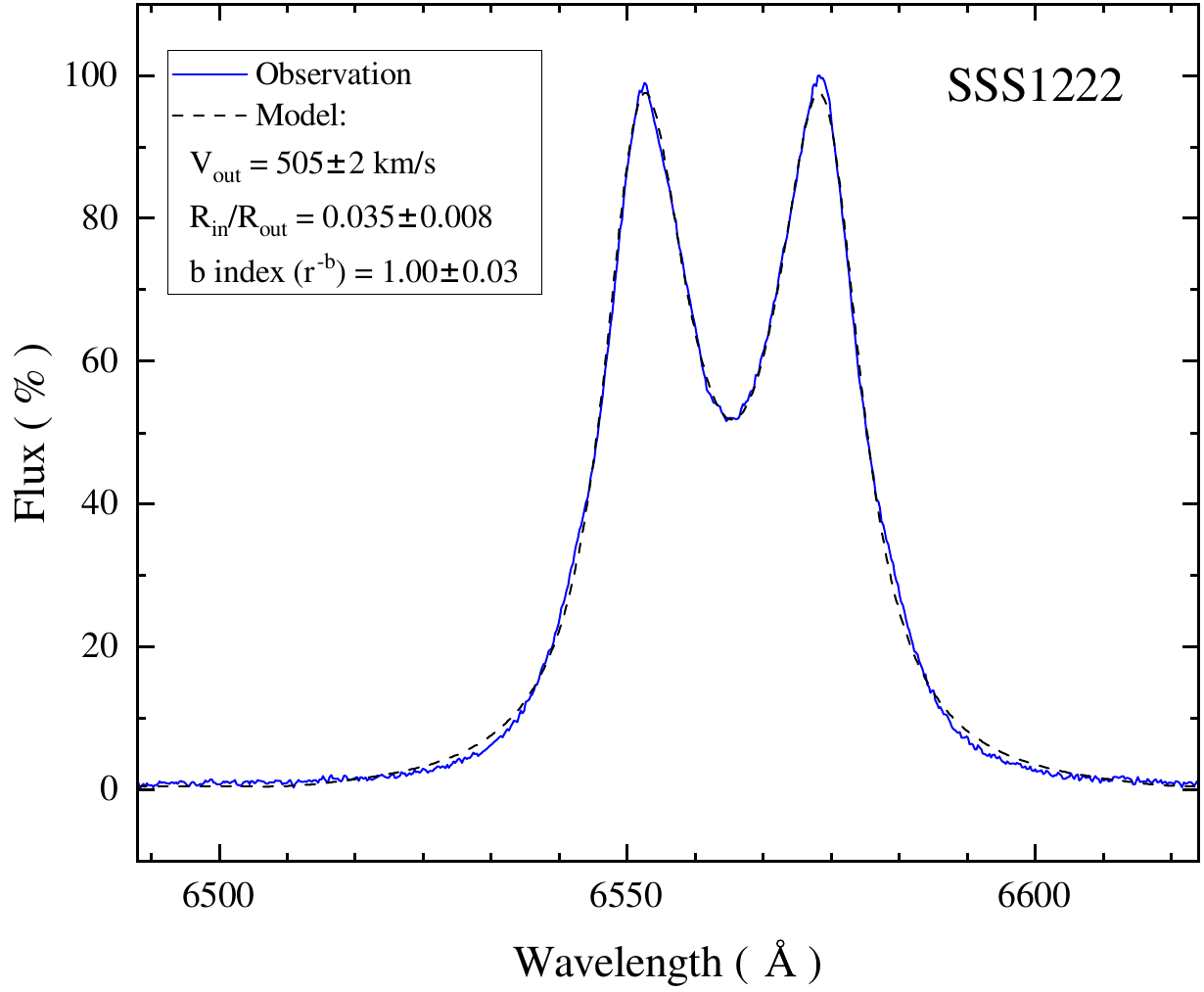} \\
    \caption{Profiles of the \Halpha\ emission lines shown together with the corresponding model fits.}
    \label{Fig:ProfileFit}
\end{figure}

\begin{table}
\begin{center}
\caption{The parameters of the accretion discs in selected WZ~Sge-type systems from the double-peaked profiles of the \Halpha\ emission line.}
\begin{tabular}{lcccc}
\hline \hline
Object  & Year of observation &$V_{\rm out}$ (\kms)&  $b$  &  $r_{\rm in}/r_{\rm out}$\\
\hline
BW Scl    & 2010 & 770$\pm$3 &  0.06$\pm$0.05 &    0.013$\pm$0.011 \\
EG Cnc    & 2019 & 659$\pm$2 &  0.71$\pm$0.03 &    0.027$\pm$0.011 \\
EZ Lyn    & 2018 & 686$\pm$2 &  0.52$\pm$0.03 &    0.032$\pm$0.009 \\
SSS J1222 & 2019 & 505$\pm$2 &  1.00$\pm$0.03 &    0.035$\pm$0.008 \\
WZ Sge    & 2025 & 780$\pm$2 &  0.18$\pm$0.03 &    0.024$\pm$0.011 \\
\hline
\end{tabular}
\label{Tab:ADparameters}
\end{center}
\end{table}

Here we are mostly interested in the radial distribution of the local line emissivity, which defines the shape of the profile
wings \citep{Smak81,HorneMarsh86,Horne95,BorisovNeustroev1998}. It is commonly assumed that the surface radial emissivity profile
follows a power-law model of the form $f(r)$ $\varpropto$ $r^{-b}$, where $r$ is the radial distance from the WD. A larger value
of $b$ makes the wings shallower. Modelling of double-peaked line profiles found in spectra of ordinary DNe and low mass X-ray
binaries has shown that $b$ is usually found to be in the range of 1--2, rarely being
less than 1.5 \citep{JohnstonKulkarniOke89,hor91ADemissionline,Orosz94,oro02j1550,neu02ippegspiral,neu16htcas}.

However, the emission line profiles in the WZ~Sge-type stars appear to be steeper, indicating a smaller $b$, and in turn,
a flatter radial emissivity profile in their discs. To estimate the $b$ and other disc parameters, we fitted the averaged
\Halpha\ profiles of the same as above  WZ~Sge-type objects, using a code implementing a simple model of a uniform flat axisymmetric
Keplerian geometrically thin disc \citep{BorisovNeustroev1998,neu02ippegspiral}. We find that $b$ in most of these objects is
notably smaller than 1.0 (see Table~\ref{Tab:ADparameters} and Figure~\ref{Fig:ProfileFit}).
This finding indicates that the disc in these systems is close to being radially isothermal, with a surface density that varies
only mildly with radial distance from the WD.

\subsection{Accretion disc and hotspot structure in quiescence}
\label{Sec:Dopmaps}

\begin{figure}
    \centering
    \includegraphics[width = 0.30\textwidth]{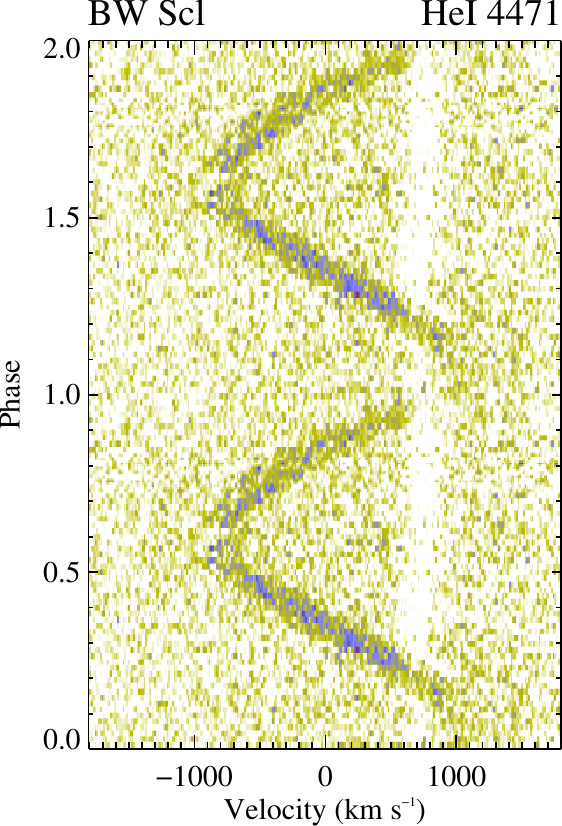}
    \includegraphics[width = 0.30\textwidth]{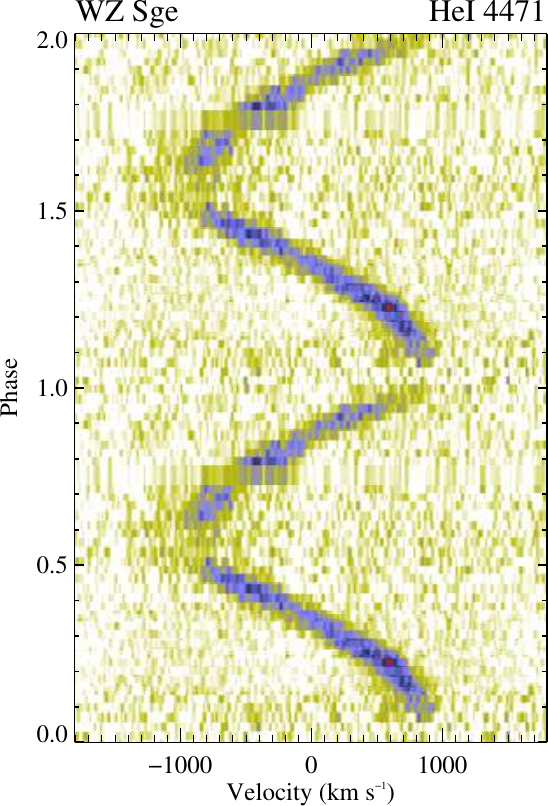}
    \includegraphics[width = 0.30\textwidth]{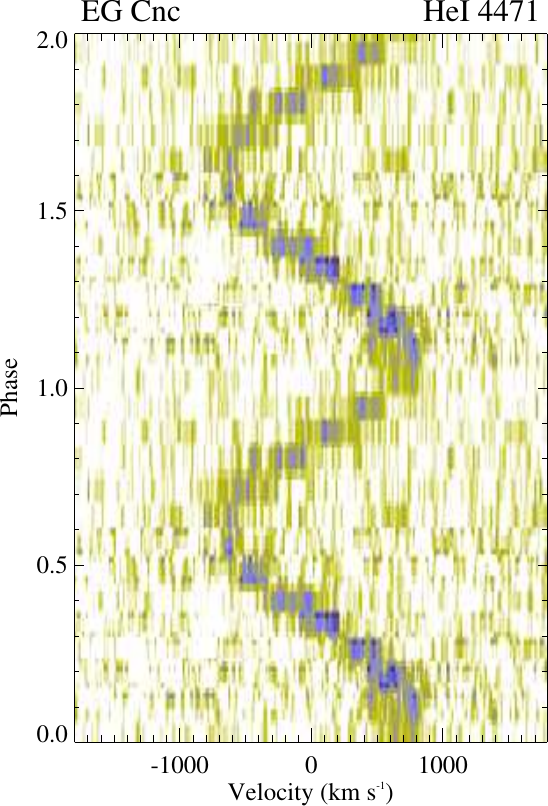} \\
    \includegraphics[width = 0.30\textwidth]{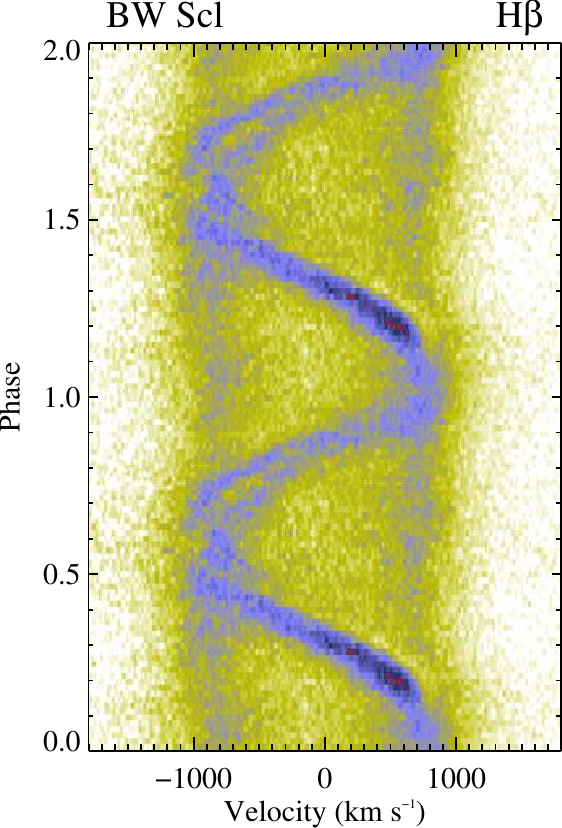}
    \includegraphics[width = 0.30\textwidth]{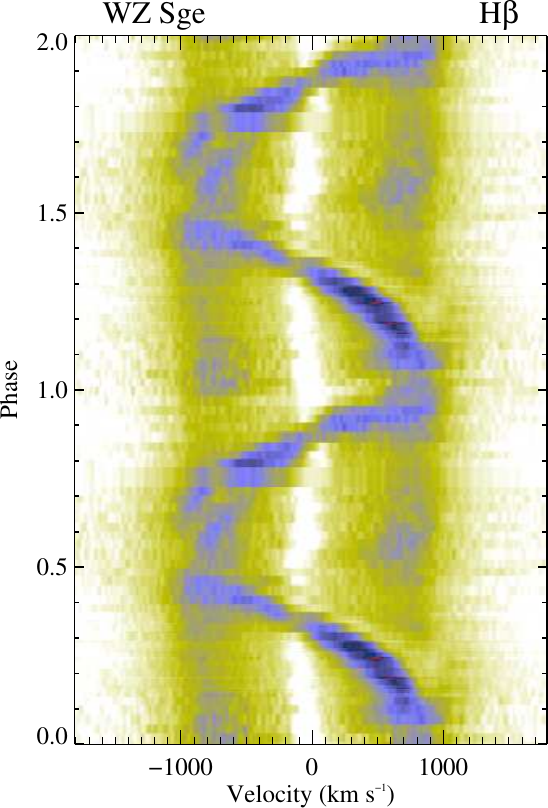}
    \includegraphics[width = 0.30\textwidth]{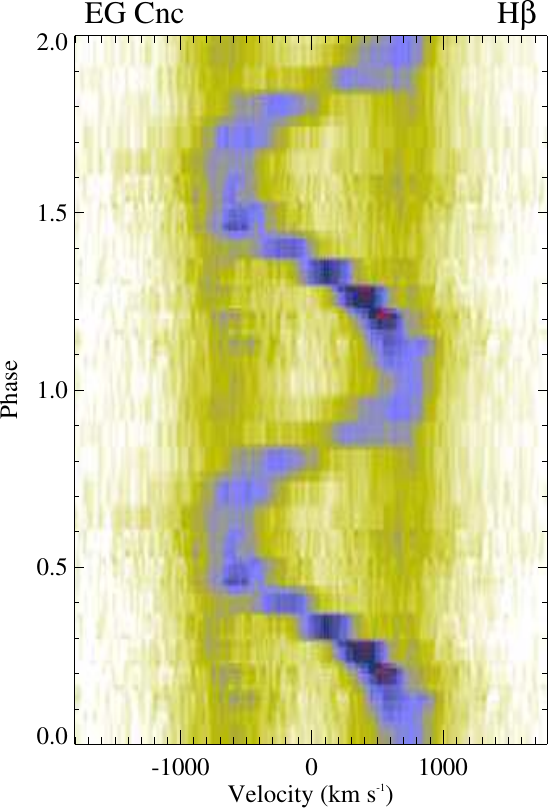}
    \caption{Trailed spectra of the \HeI\ 4471\AA\ (upper panels) and \Hbeta\ (bottom panels) emission lines in BW~Scl (left), WZ~Sge
    (middle), and EG~Cnc (right). In all these objects, the helium lines are dominated by the hotspot, while the accretion disc
    contribution is negligible. The Balmer lines, however, show a mixture of emission components from both the disc and hotspot.
    The hotspot S-wave of the Balmer lines exhibits a complex behaviour, splitting when it is blue-shifted, and displaying
    an absorption shadow when it is red-shifted. See \cite{neu23bwscl} for more detail.
    }
    \label{Fig:TrailedSpec}
\end{figure}

\begin{figure}
    \centering
    \includegraphics[width = 0.32\textwidth]{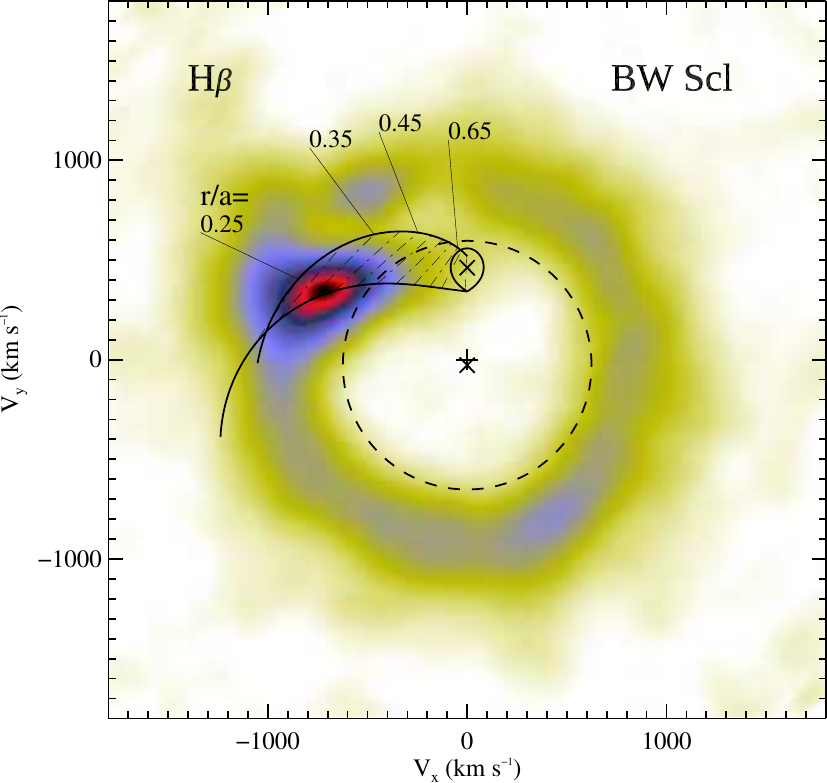}
    \includegraphics[width = 0.32\textwidth]{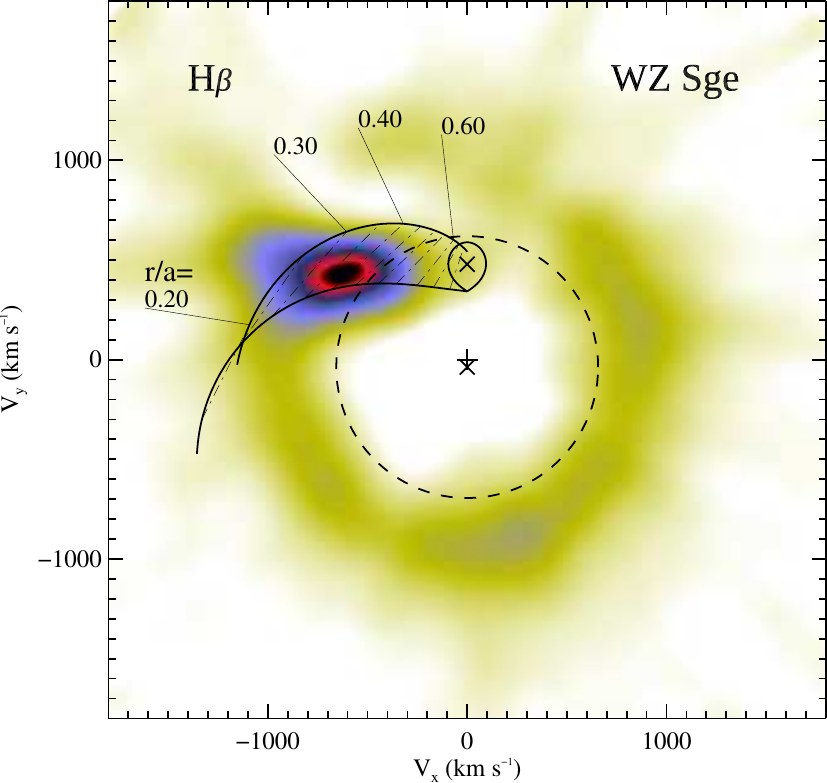}
    \includegraphics[width = 0.32\textwidth]{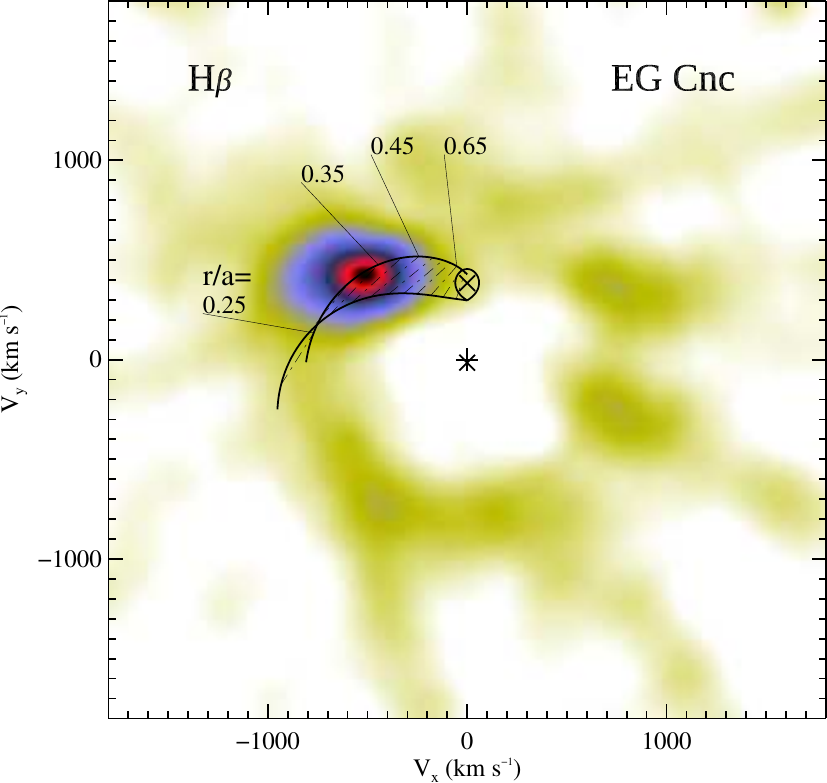} \\
    \includegraphics[width = 0.32\textwidth]{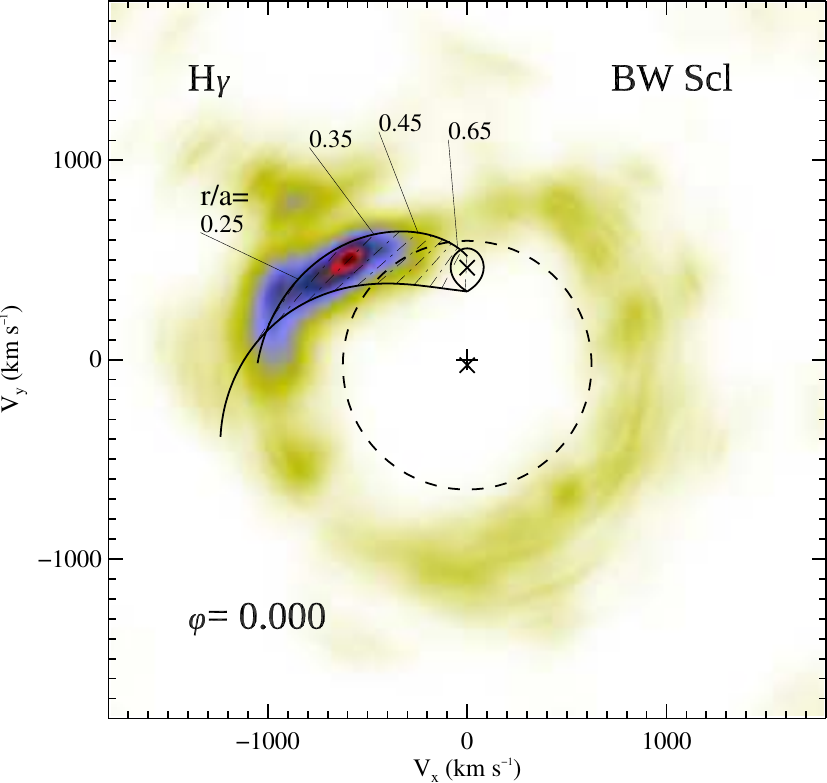}
    \includegraphics[width = 0.32\textwidth]{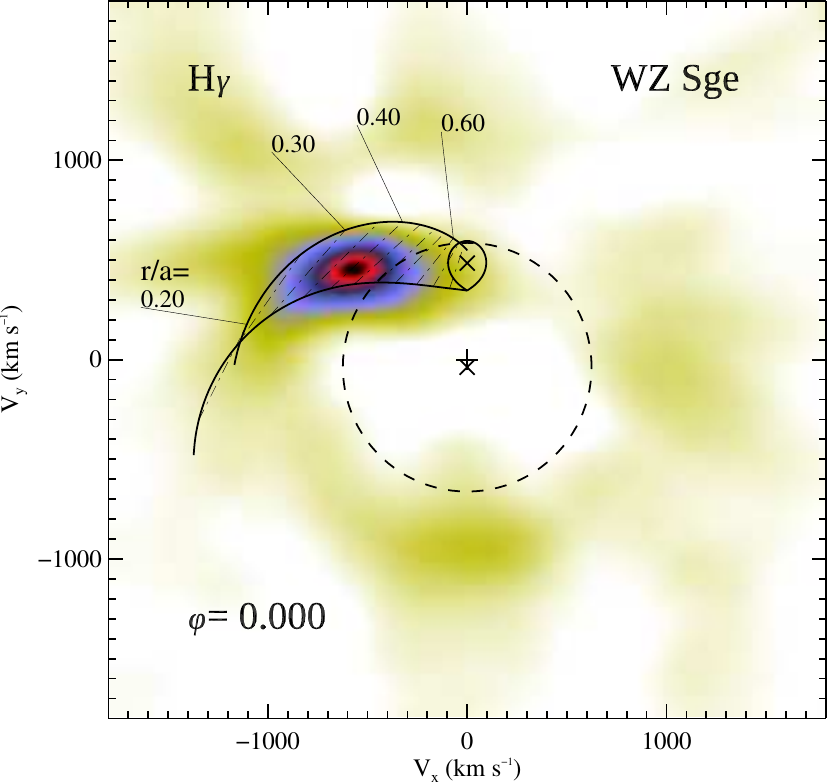}
    \includegraphics[width = 0.32\textwidth]{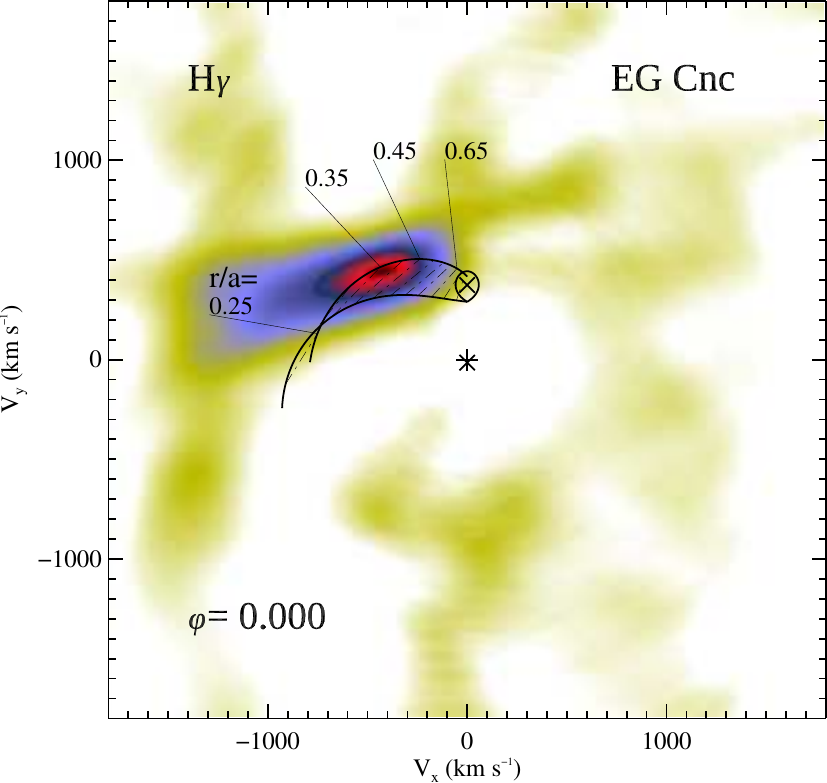} \\
    \includegraphics[width = 0.32\textwidth]{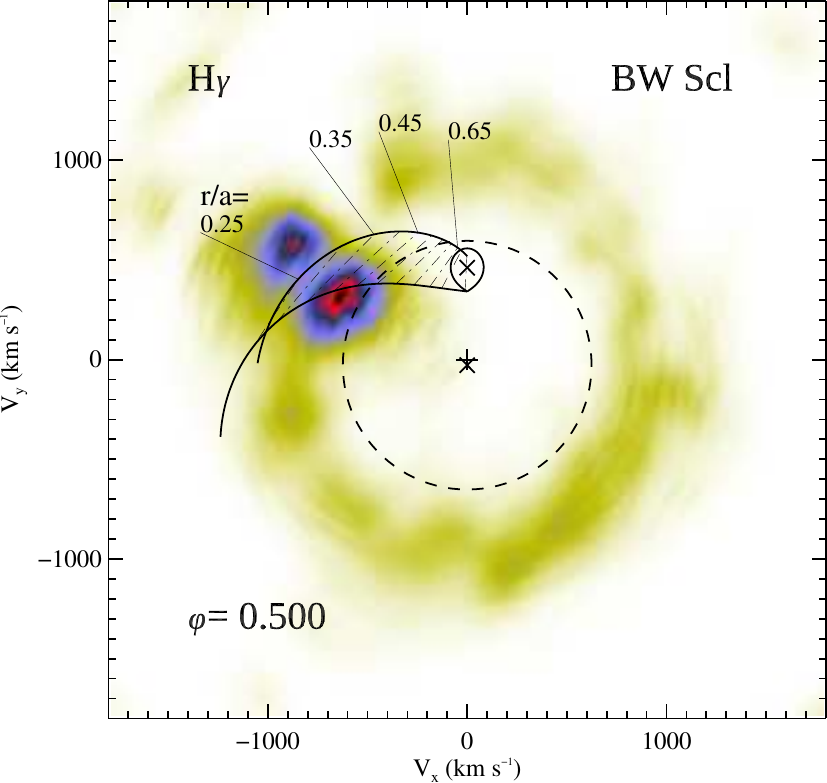}
    \includegraphics[width = 0.32\textwidth]{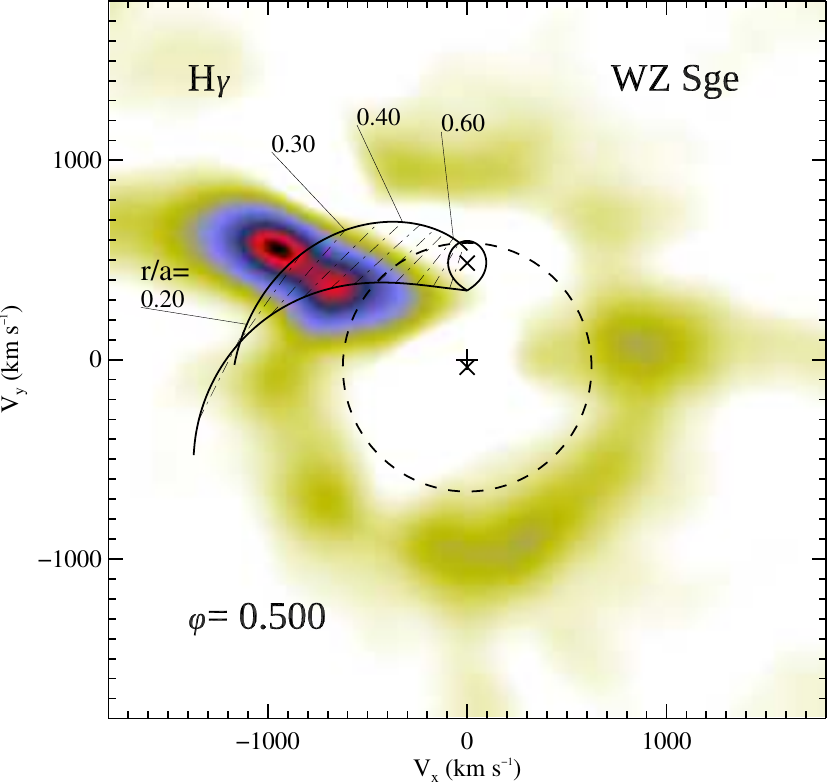}
    \includegraphics[width = 0.32\textwidth]{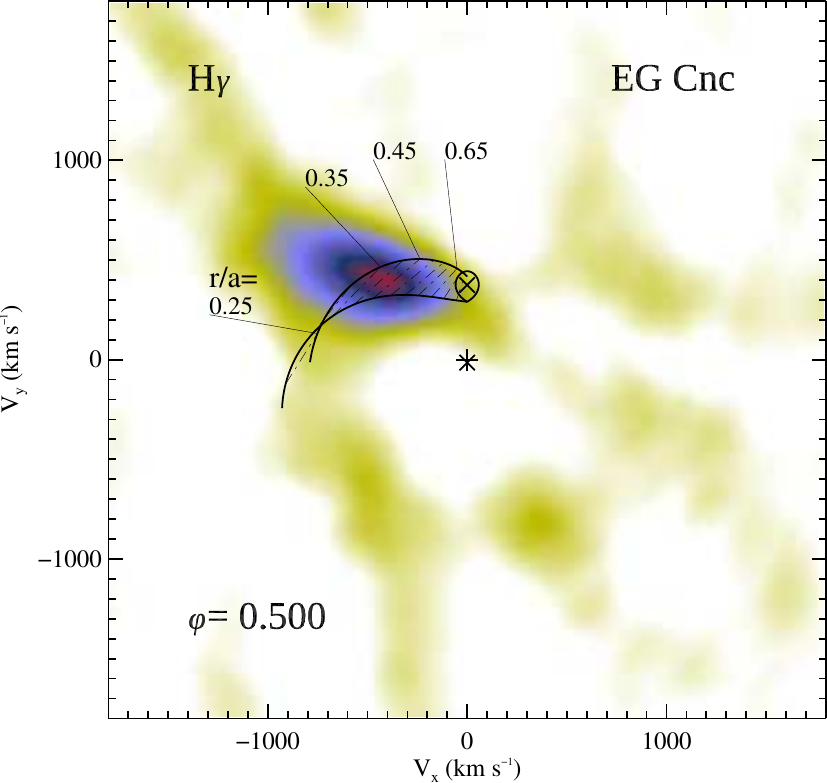}
    \caption{Doppler maps of BW~Scl (left column, WZ~Sge (middle column), and EG~Cnc (right column). The upper maps of
    the \Hbeta\ emission line are calculated using the whole sets of spectra, whereas for the lower maps of the \Hgamma\
    line, only 50 per cent of spectra between phases 0.75--1.25 (middle panels) and 0.25--0.75 (bottom panels) were used.
    The dashed circles show the tidal truncation radius.
    The dashed lines connect the velocity of the ballistic gas stream (lower curve) and the velocity on the Keplerian
    disc along the gas stream (upper curve) for the same points at distances labelled along the upper curve (in r/a units).
    These lines are separated by 0.05$r$/$a$. See \cite{neu23bwscl} for more detail.
    }
    \label{Fig:Dopmaps-total}
\end{figure}

High time- and spectral-resolution data for a sample of WZ Sge-type objects enabled us to establish some properties of accretion
discs that are particularly important in the context of this paper. These properties seem consistent across the entire subclass.
\begin{enumerate}
 \item From the trailed spectra one can see (Figure~\ref{Fig:TrailedSpec}) that the accretion disc doesn't produce any helium
     emission as all the observed \HeI\ and \HeII\ lines come primarily from the hotspot (see \cite{neu23bwscl} for more detail).
     Thus, the disc is not hot enough to excite any helium and even higher-order Balmer emission lines.
 \item Doppler maps of objects with known accurate system parameters (e.g., BW Scl and WZ~Sge -- \cite{neu23bwscl,KvistDIM50},
     see also the ordinary SU UMa-type star HT Cas -- \cite{neu16htcas,neu20HTCasoutburstdopmap}) show that the accretion disc,
     even in quiescence, is large, reaching the truncation radius (Figure~\ref{Fig:Dopmaps-total}), which is larger than the 3:1
     resonance radius.
 \item The hotspot has a complex structure, and when Dynamical Doppler tomography\footnote{Dynamical tomograms for BW~Scl
     are available at \url{https://vitaly.neustroev.net/researchfiles/bwscl/}} is applied (see \cite{neu23bwscl} for more detail),
     it is seen that the hotspot is highly anisotropic. Two bottom rows of panels in Figure~\ref{Fig:Dopmaps-total} show the
     Doppler maps calculated using 50 per cent of spectra centred on phases 0.0 (0.75--1.25) and 0.5 (0.25--0.75).
 \item The Doppler maps show that the outer parts of the accretion disc have low density, allowing the gas stream to
     flow along a ballistic trajectory into the inner disc regions.
 \item The elongated hotspot becomes visible at or even beyond the disc truncation radius, but its brightest part is located
     close to the circularisation radius of the disc \citep{neu23bwscl}.
\end{enumerate}

\begin{figure}
    \centering
    \includegraphics[width = 0.49\textwidth]{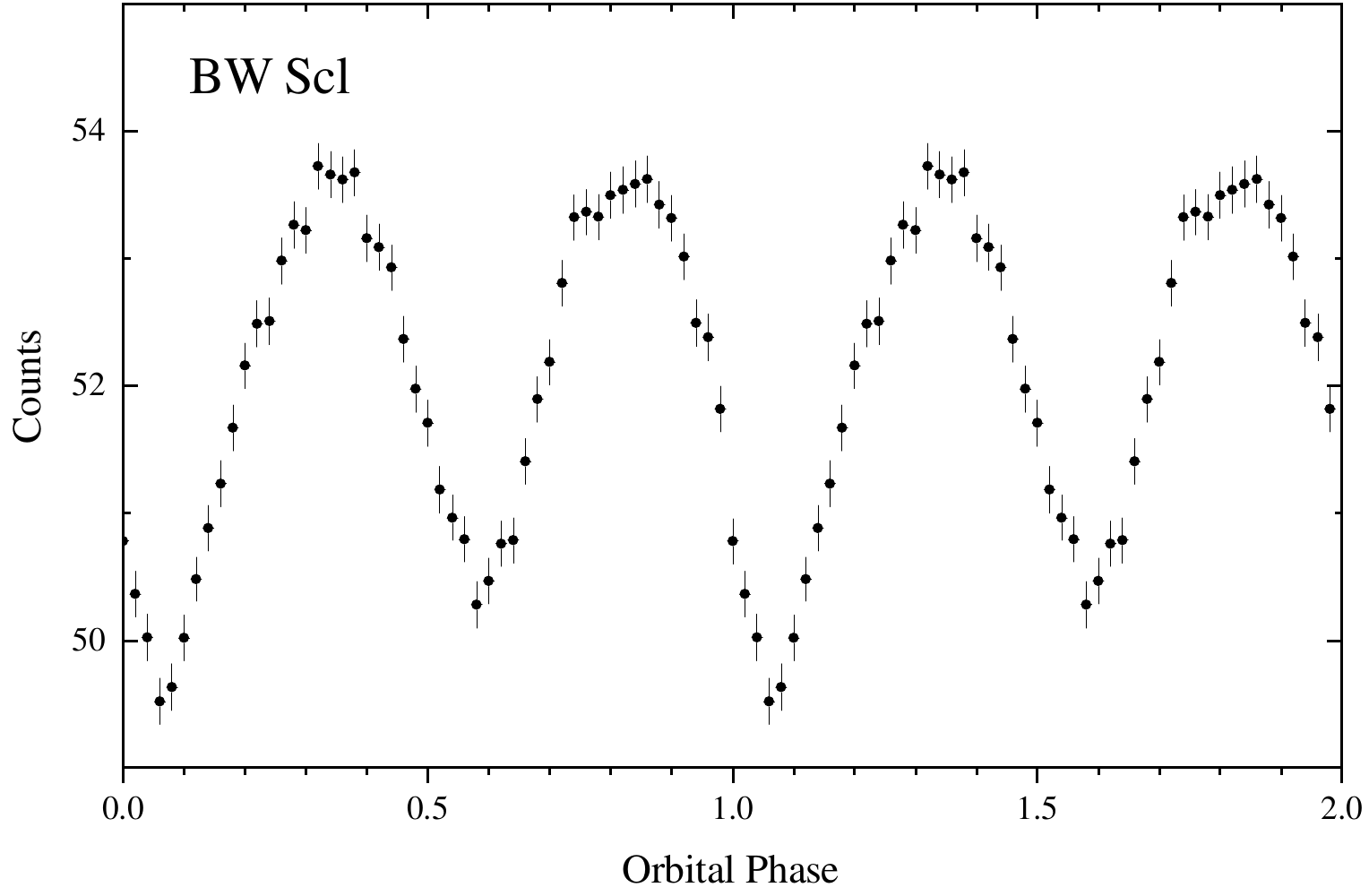}
    \includegraphics[width = 0.49\textwidth]{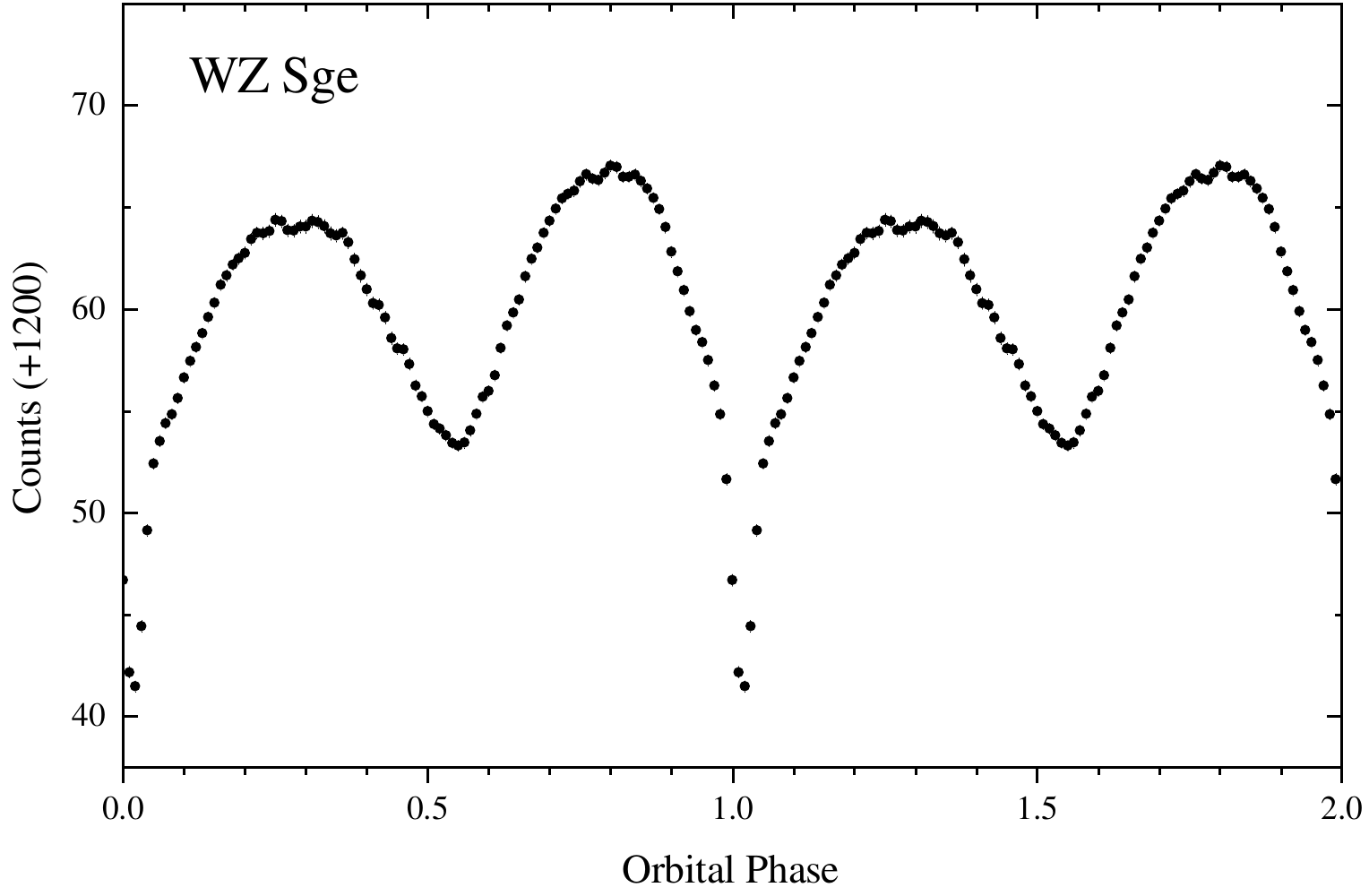}
    \caption{The TESS light curves of BW Scl (left) and WZ Sge (right) folded with their orbital periods and averaged in
    50 phase bins. WZ Sge is an eclipsing CV showing, in addition to double-wave modulations, eclipses of the hotspot on
    the disc rim.}
    \label{Fig:TESS}
\end{figure}

Periodograms of optical light curves of WZ~Sge-type stars are dominated by a coherent signal at half the orbital frequency,
so-called double-wave modulations (Figure~\ref{Fig:TESS}). Such photometric behaviour is naturally explained by an elongated,
optically thick hotspot (whose viewing aspect varies with the orbital period) shining through/above an optically thin accretion
disc. An almost transparent disc, as seen in these objects, allows the light from the elongated hotspot to escape in all
directions, while the variable aspect of the hotspot modulates the observed flux.

\begin{figure}
    \centering
    \includegraphics[width = 0.5\textwidth]{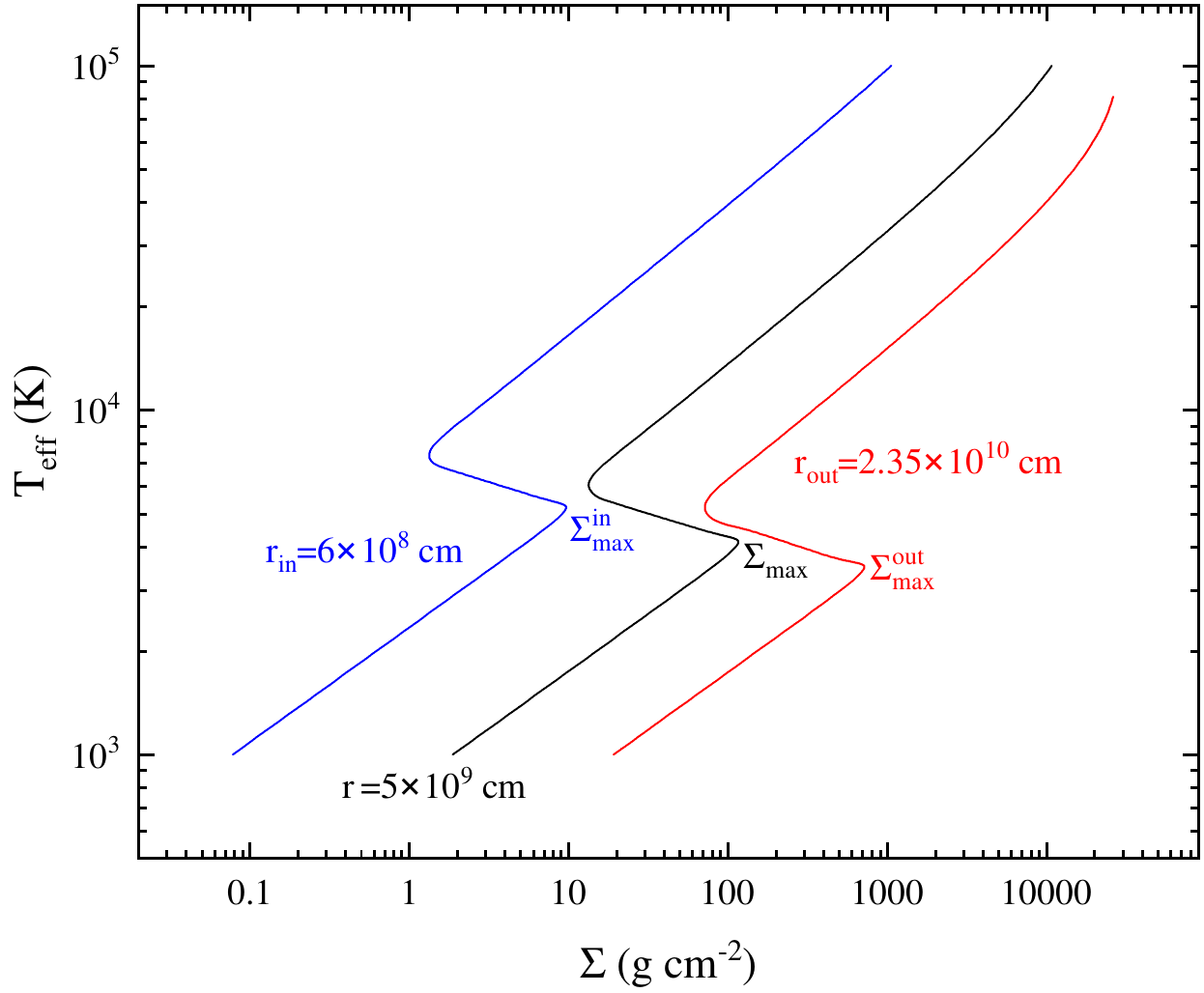}
    \caption{Example of $\Sigma$ -- $T_{\rm eff}$ S-curves calculated for the mass of BW~Scl $M_{\rm wd}$=0.85 \Msun\ at its
    inner (6\tim{8} cm) and outer (2.35\tim{10} cm) disc radii, and also at 5\tim{9} cm. The cold and hot branches of the the
    S-curves were calculated with $\alpha$ = 0.01 and 0.1 respectively, using the code from  \cite{Tavleev+23}.}
    \label{Fig:Scurve}
\end{figure}

\section{Predictions of the DIM and comparison with observations}

According to the DIM, the quiescent disc in a DN just before its outburst must be filled, at some radius, to the critical surface
density $\Sigma_{\rm max}$ which corresponds to the turning point of the lower stable branch of the S-curve, connecting it to the
intermediate, thermally and viscously unstable, branch (Figure~\ref{Fig:Scurve}).\footnote{$\Sigma_{\rm max}$ is the maximum
surface density the accretion disc can sustain in a stable, cold quiescent state. If the surface density $\Sigma$ locally exceeds
this critical value, the disc becomes thermally unstable and transitions to a hot, high-viscosity state, leading to an outburst.}
$\Sigma_{\rm max}$ depends on the viscosity parameter $\alpha$, the mass of the WD, and the radial distance from the centre. This
parameter can be estimated directly from the S-curve, but can also be calculated using one of the published fits to
$\Sigma_{\rm max}$ (e.g., eqns. A.1 in \cite{las01DIDNXT}, 16 in \cite{ham20CVreview}, A3 in \cite{Bollimpalli+18}). Thus, the
smallest value of $\Sigma_{\rm max}$ is expected to be found at the inner disc radius, although both the observations and the TTI
model support the outside-in outbursts in the WZ~Sge-type DNe. For these two cases, the value of $\Sigma_{\rm max}$ in BW~Scl for
adopted $\alpha$=0.01 ranges between $\sim$10 g cm$^{-2}$ at the inner disc radius to $\sim$1000 g cm$^{-2}$ at the outer one,
and it is larger for smaller $\alpha$ ($\Sigma_{\rm max}$ $\propto$ $\alpha^{-0.83}$ \cite{ham20CVreview}).
As even the smallest $\Sigma_{\rm max}$ value corresponds to optically thick conditions, it indicates that the whole disc at the
onset of the outburst must be optically thick. This is, in fact, one of the defining assumptions of the DIM \cite{las01DIDNXT}.

However, we found multiple lines of observational evidence that the entire accretion discs in WZ~Sge-type stars in deep quiescence
are optically thin. They have a very low bolometric luminosity (a few $\times$10$^{30}$ erg s$^{-1}$ which corresponds to a very
low-mass accretion rate of a few $\times$$10^{-13}$ \Msun\ yr$^{-1}$). An order-of-magnitude estimation of the average surface
density of the disc gives $\overline{\Sigma}$ $\ll$ $\Sigma_{\rm max}$. These conditions within the disc cannot trigger an outburst.

\begin{figure}
    \centering
    \includegraphics[width = 0.49\textwidth]{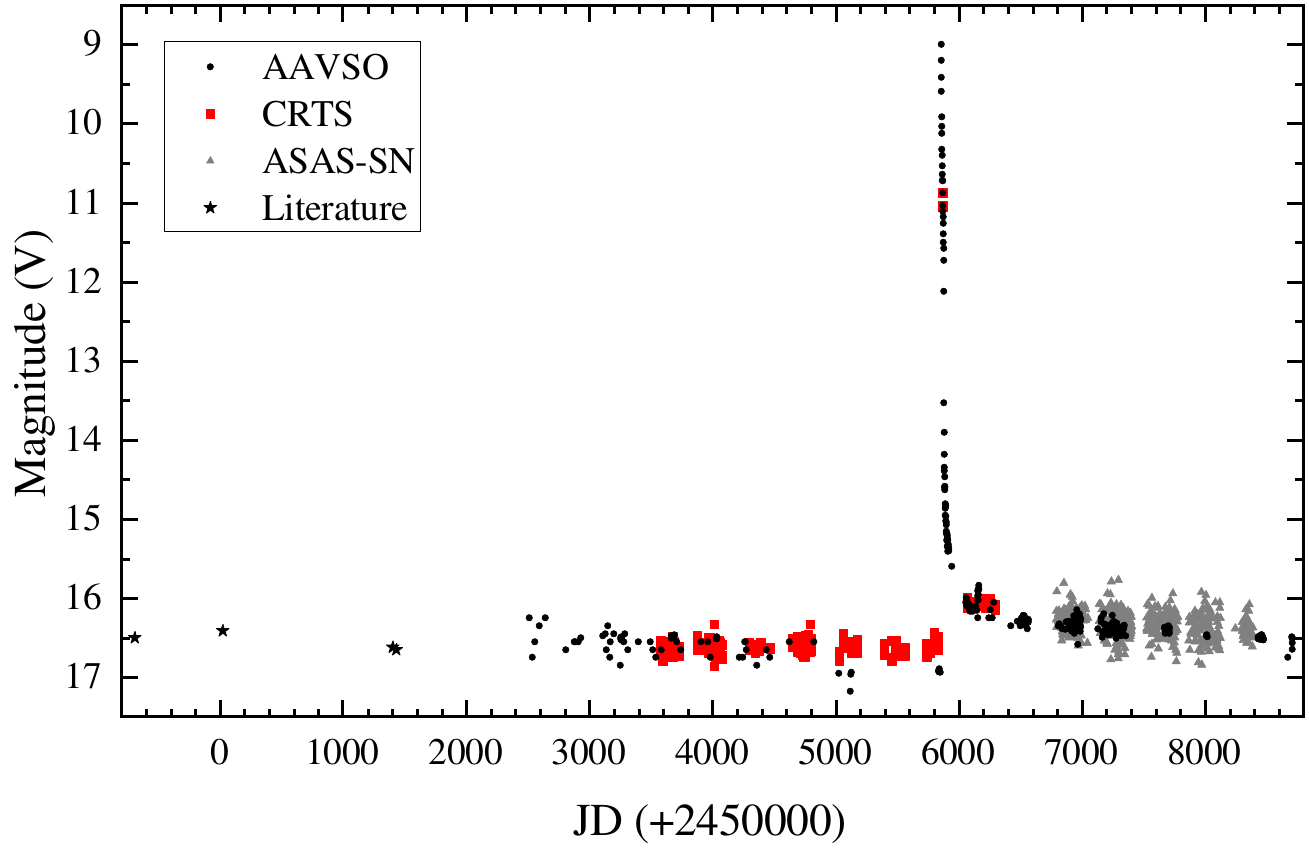}
    \includegraphics[width = 0.49\textwidth]{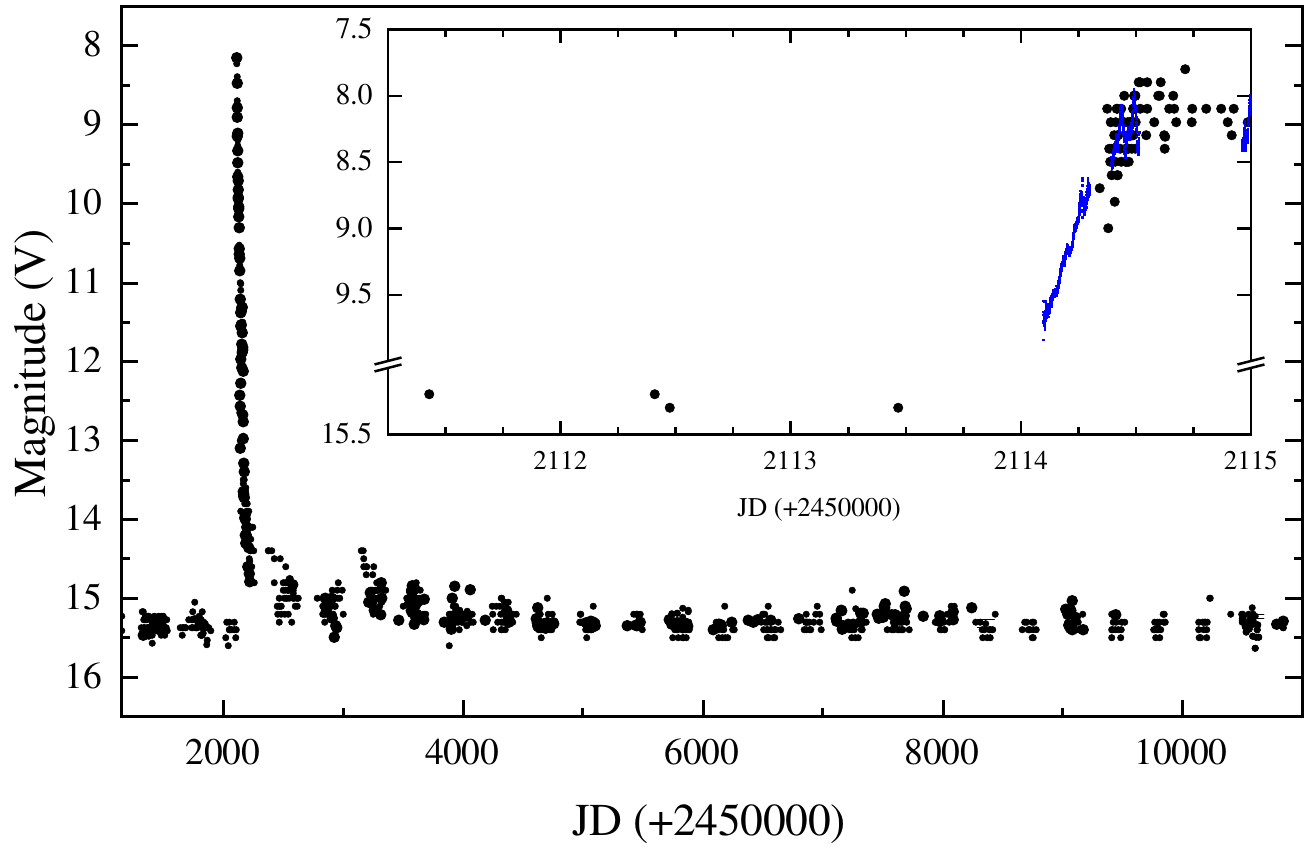}
    \caption{Long-term light curves of BW~Scl (left-hand panel, see \cite{neu23bwscl} for more detail) and WZ~Sge
    (right-hand panel). The inset plot in the right-hand panel shows an enlarged region around the rising stage of
    the superoutburst. The blue data points represent observations by Ishioka et al. \cite{ish02wzsgeletter}.
    }
    \label{Fig:LongTermLC}
\end{figure}

The pre-outburst state of DNe has been considered multiple times in the past
\cite{mey81DNoutburst,can82DNburst,sma82alphadisk,mey88DNpreoutburst,MeyerHofmeisterRitter93}, though without comprehensive
observational support. Meyer in \cite{mey88DNpreoutburst} (see also \cite{MeyerHofmeisterRitter93}) has explicitly shown that
though the disc becomes totally optically thick almost simultaneously with the start of an outburst, the disc becomes gradually
thick more than two weeks before the onset of instability, for the considered $\dot{M}_{\rm acc}$=3\tim{-10}~\Msun\ yr$^{-1}$
(figure 4 in \cite{mey88DNpreoutburst}). One can expect that this time interval will be much more prolonged for significantly
lower $\dot{M}_{\rm acc}$ estimated in WZ~Sge-type stars.

As was discussed above, the optical flux difference between a totally optically thin and even a partially optically thick disc
should be more than 2 magnitudes. Such an increase in brightness would be easily detected even by amateur astronomers. However,
the pre-outburst photometry available for a few WZ~Sge-type objects does not show any trend in brightness
(Figure~\ref{Fig:LongTermLC}). For example, the prototype object WZ Sge had the same average brightness for more than 10 years,
up to half a day before the 2001 superoutburst (the inset plot in the right-hand panel of Figure~\ref{Fig:LongTermLC}),
the same as we observe it now. Now, 24 years after the previous superoutburst,
the accretion disc in WZ Sge has a very low luminosity ($\lesssim$1.6$\times$10$^{30}$ erg s$^{-1}$), indicating a very low
mass-transfer rate ($\sim$1.0$\times$$10^{13}$ g s$^{-1}$). We see no significant change in the accretion disc's overall
brightness compared with this or the previous quiescent cycle. If this cycle follows the previous one, WZ~Sge should outburst
soon. There is no sign of the accretion disc becoming optically thick (for more detail on the long-term evolution of WZ~Sge
throughout quiescence see \cite{KvistDIM50}).

To explain the extraordinary behaviour of WZ Sge-type systems, the DIM requires
\cite{ham97wzsgemodel,bua02suumamodel,ham21V3101Cygrebrightening} that the inner disc be sufficiently truncated\footnote{The
truncation prevents the inner accretion disc from becoming thermally unstable, forcing material to accumulate in the outer,
cooler parts of the disc where $\Sigma_{\rm max}$ is much larger than at lower radii.} and that a superoutburst be triggered
by an enhancement of mass-transfer from the donor star. This increase of $\dot{M}_{\rm tr}$ must be very significant, by several
orders of magnitude \cite{ham21V3101Cygrebrightening}. However, even without considering the reason for this increase or why it
occurs regularly (e.g., in WZ~Sge), this explanation has serious drawbacks. Even a very substantial rise in $\dot{M}_{\rm tr}$
prior to the outbursts will not immediately trigger a thermal and a viscous instability, as the low-density disc needs to
accumulate enough material for the outburst. This will require at least days, or even weeks, during which the disc will first
transition to a partially optically thick regime, which is readily observable, as shown above. Moreover, a sudden increase of
$\dot{M}_{\rm tr}$ will proportionally brighten the hotspot. Even a modest 10-fold increase in hotspot luminosity will result
in an average system brightening of 0.5-1 mag and extremely strong orbital modulations, which will still be seen during the
rising stage of the outburst. Nothing of it was observed in WZ~Sge (half a day before the 2001 superoutburst detection, it had
the same brightness as before; see the inset plot in the right-hand panel of Figure~\ref{Fig:LongTermLC}) or in a few other
similar objects.

Another hypothesis about disc truncation is also not supported by our observations. From the fit of double-peaked profiles of
the \Halpha\ emission line, we obtained the ratios of the disc inner and outer radii for several WZ Sge-type objects. On average,
$r_{\rm in}/r_{\rm out}$$\approx$0.03$\pm$0.01 which is in agreement with the estimated values of $R_{\rm wd}/r_{\rm out}$. For
example, in BW~Scl it was found to be $R_{\rm wd}/r_{\rm out}$=0.030 \cite{neu23bwscl}. Adopting for $r_{\rm in}$ of the
truncated disc to be 4\t$R_{\rm wd}$, we obtain $r_{\rm in}/r_{\rm out}$=0.12. We note that the wings of double-peaked profiles
are very sensitive to $r_{\rm in}/r_{\rm out}$ (see, e.g. figure 2 in \cite{BorisovNeustroev1998}), and with such a value it will
be definitely detected.

\section{Conclusion and open questions}

We presented an analysis of observations of a sample of WZ Sge-type systems in deep quiescence, aimed at assessing how
the predictions of the disc instability model (DIM) compare with their observed properties. In particular, we investigated
whether superoutbursts in WZ Sge-type systems are indeed triggered by enhanced mass transfer from the donor star. By applying
a consistent methodology to all objects in the sample, we obtained comparable and robust results. We found that the accretion
discs in quiescent WZ Sge-type systems exhibit very low bolometric luminosities of a few $\times 10^{30}$ erg s$^{-1}$,
corresponding to extremely low mass-accretion rates of (2$\div$3)\tim{-13} \Msun\ yr$^{-1}$. Such discs are entirely optically
thin. In this regime, the physical conditions within the disc, such as surface density and effective temperature, remain well
below the DIM thresholds and therefore cannot trigger a superoutburst. Moreover, observationally, the disc brightness does not
change prior to the superoutburst, confirming that there is no evidence for a transition to an optically thick regime, even
though the DIM predicts that the disc should become progressively thicker for at least several days, and most likely weeks,
before the onset of the instability.

Thus, we conclude that there is no evidence that superoutbursts in WZ Sge-type systems are triggered by an enhanced
mass-transfer rate. Consequently, the fundamental question remains open: how do accretion discs in these systems enter outburst?
Still, we admit that the observations of WZ~Sge-type objects just prior to their superoutbursts are very limited. Multi-colour
photometric and especially spectroscopic observations are critically needed to shed more light on the pre-outburst behaviour of
these objects.

\acknowledgments
We thank the Fujihara Foundation of Science for financial support that made our participation in this workshop possible.
We acknowledge funding to support the scientific use of our ESO/ALMA data from the Finnish Centre for Astronomy with ESO (FINCA),
University of Turku, Finland.

Based on observations made with ESO Telescopes at the La Silla Paranal Observatory under programme IDs 086.D-0775, 100.D-0932,
101.D-0806, 103.D-0110, and 115.D-0144, and with the Nordic Optical Telescope, operated by the Nordic Optical Telescope
Scientific Association at the Observatorio del Roque de los Muchachos, La Palma, Spain, of the Instituto
de Astrofisica de Canarias. The data presented here were obtained in part with ALFOSC, which is provided
by the Instituto de Astrofisica de Andalucia (IAA) under a joint agreement with the University of
Copenhagen and NOTSA. This paper includes data collected by the \textit{TESS} mission, which are publicly
available from the Mikulski Archive for Space Telescopes (MAST). Funding for the \textit{TESS} mission is provided by the
NASA's Science Mission Directorate. This publication makes use of data products from the \textit{Wide-field Infrared Survey
Explorer}, which is a joint project of the University of California, Los Angeles, and the Jet Propulsion Laboratory/California
Institute of Technology, funded by the National Aeronautics and Space Administration. 
We thank the {\it Swift} PI, Brad Cenko, for approving the observations, and the {\it Swift} planning and operations teams for
their support. We acknowledge with thanks the variable star observations from the {\it AAVSO International Database} contributed
by observers worldwide and used in this research.

\bibliographystyle{JHEP}

\end{document}